\documentclass[journal, twocolumn]{IEEEtran}

\usepackage{cite}
\usepackage{graphicx}
\usepackage{amsmath}
\usepackage{caption}
\usepackage{booktabs}
\usepackage{subfigure}

\usepackage[norelsize, linesnumbered, ruled, lined, boxed, commentsnumbered]{algorithm2e}      
\SetKwRepeat{Do}{do}{while}%
\SetKw{Kwin}{ in } 
\SetKw{KwTo}{ to }
\SetKw{KwGoto}{goto }

\usepackage{setspace}

\newcommand{\EquRef}[1]{Eq. \ref{#1}}
\newcommand{\SecRef}[1]{Sec. \ref{#1}}
\newcommand{\SubSecRef}[1]{Sec. \ref{#1}}
\newcommand{\TableRef}[1]{Tab. \ref{#1}}
\newcommand{\FigRef}[1]{Fig. \ref{#1}}
\newcommand{\Athr}[1]{{#1} \emph{et al.}}
\newcommand{\AlgRef}[1]{Alg. \ref{#1}}

\usepackage{amssymb}   
\usepackage[hidelinks]{hyperref}
\begin{document}


\title{PPO-Based Vehicle Control for Ramp Merging Scheme Assisted by Enhanced C-V2X}

\author{
    Qiong Wu, ~\IEEEmembership{Senior Member,~IEEE}, Maoxin Ji, Pingyi Fan, ~\IEEEmembership{Senior Member,~IEEE},\\ Kezhi Wang, ~\IEEEmembership{Senior Member,~IEEE}, Nan Cheng, ~\IEEEmembership{Senior Member,~IEEE},\\ Wen Chen, ~\IEEEmembership{Senior Member,~IEEE}, and Khaled B. Letaief, ~\IEEEmembership{Fellow,~IEEE}

    \thanks{
    	Part of this work was presented in International Conference on Communication Technology (ICCT), 20-22 Oct. 2023, Wuxi, China \cite{10419638}.
    	
        Qiong Wu and Maoxin Ji are with the School of Internet of Things Engineering, Jiangnan University, Wuxi 214122, China (e-mail: qiongwu@jiangnan.edu.cn, maoxinji@stu.jiangnan.edu.cn).
        
        Pingyi Fan is with the Department of Electronic Engineering, State Key laboratory of Space Network and Communications, Beijing National Research Center for Information Science and Technology, Tsinghua University, Beijing 100084, China (e-mail: fpy@tsinghua.edu.cn).
        
        Kezhi Wang is with the Department of Computer Science, Brunel University, London, Middlesex UB8 3PH, U.K (e-mail: Kezhi.Wang@brunel.ac.uk).
        
        Nan Cheng is with the State Key Lab. of ISN and School of Telecommunications Engineering, Xidian University, Xi'an 710071, China (e-mail: dr.nan.cheng@ieee.org).
        
        Wen Chen is with the Department of Electronic Engineering, Shanghai Jiao Tong University, Shanghai 200240, China (e-mail: wenchen@sjtu.edu.cn).
        
        Khaled B. Letaief is with the Department of Electrical and Computer Engineering, the Hong Kong University of Science and Technology (HKUST), Hong Kong (email: eekhaled@ust.hk).
        }
}
\maketitle

\begin{abstract}

On-ramp merging presents a critical challenge in autonomous driving, as vehicles from merging lanes need to dynamically adjust their positions and speeds while monitoring traffic on the main road to prevent collisions. To address this challenge, we propose a novel merging control scheme based on reinforcement learning, which integrates lateral control mechanisms. This approach ensures the smooth integration of vehicles from the merging lane onto the main road, optimizing both fuel efficiency and passenger comfort. Furthermore, we recognize the impact of vehicle-to-vehicle (V2V) communication on control strategies and introduce an enhanced protocol leveraging Cellular Vehicle-to-Everything (C-V2X) Mode 4. This protocol aims to reduce the Age of Information (AoI) and improve communication reliability. In our simulations, we employ two AoI-based metrics to rigorously assess the protocol's effectiveness in autonomous driving scenarios. By combining the NS3 network simulator with Python, we simulate V2V communication and vehicle control simultaneously. The results demonstrate that the enhanced C-V2X Mode 4 outperforms the standard version, while the proposed control scheme ensures safe and reliable vehicle operation during on-ramp merging.
	
\end{abstract}

\IEEEpeerreviewmaketitle

\section{Introduction}\label{intro}

\IEEEPARstart{A}{ccelerated} by recent advancements in wireless communication and machine learning (ML) technologies, the development and adoption of the Internet of Vehicles (IoV) and autonomous driving systems have significantly progressed \cite{10163760, 10387423, 10032660, 10620366}. These technologies present substantial opportunities for transforming transportation systems, thereby enhancing their safety, efficiency, and intelligence \cite{10274112, schwarting2018planning}. IoV technology enables seamless communication between vehicles and various sensors, thereby significantly augmenting the capabilities of autonomous driving systems \cite{gupta2021internet, 8907851, 10402048}. Autonomous vehicles, due to their ability to make unbiased control decisions, possess the potential to drastically reduce the incidence of traffic accidents in complex traffic scenarios\cite{chen2018examining}.

Ramp merging presents a common yet challenging traffic scenario, particularly prone to traffic accidents. Statistics indicate that approximately 30$\%$ of traffic accidents in China occur during merging operations \cite{wang2013automated}. This statistic underscores the critical need for effective autonomous driving solutions that can guarantee safe on-ramp mergings. In these scenarios, vehicles entering the main road are required to determine the optimal longitudinal position and merge smoothly before the acceleration lane ends. This process demands delicate coordination to prevent collisions, requiring sophisticated control of both longitudinal and lateral movements to ensure safety and passenger comfort. The complexity of these control challenges poses a significant obstacle for autonomous on-ramp merging.

Vehicle sensors, such as cameras, radar, and LiDAR, can frequently be obstructed by roadside structures or vegetation, which may hinder the vehicle's ability to promptly gather information about other road users \cite{test-cite, 9877926}. This limitation can negatively impact decision-making quality, thereby increasing the risk of collisions. Consequently, vehicle-to-vehicle (V2V) communication is indispensable for exchanging positional and environmental data, thereby mitigating these risks.

The Internet of Vehicles (IoV) plays an important role in supporting vehicle control during on-ramp merging scenarios \cite{10234718}. In its 14th edition, the 3rd Generation Partnership Project (3GPP) introduced the Cellular Vehicle-to-Everything (C-V2X) standard to facilitate IoV communications. This standard includes several communication modes, such as Vehicle-to-Infrastructure (V2I), Vehicle-to-Device (V2D), and V2V, enabling direct vehicular communication without relying on cellular networks. C-V2X communication, known for its low latency, is particularly important for autonomous driving applications, as it allows for coordinated control and improves traffic flow.

Within the C-V2X framework, resource allocation is managed by Mode 3 and Mode 4, which respectively support centralized and decentralized operations. While Mode 3 offers superior performance through base station-assisted resource allocation, its utility is limited by coverage constraints. In contrast, C-V2X Mode 4 operates independently of base stations, making it suitable for IoT applications but susceptible to latency and packet loss due to its decentralized structure. These challenges require careful consideration of resource allocation strategies to maintain the reliability of vehicle control \cite{8672604}.

The sensing-based semi-persistent schedule (SB-SPS) in C-V2X Mode 4 has some known drawbacks. Vehicles reserving overlapping resources may experience communication failures due to half-duplex operations, which affects the Age of Information (AoI). Additionally, a lack of awareness of resource occupancy leads to transmission conflicts and lower reliability. These issues are especially critical in autonomous driving, where outdated information may result in erroneous decisions and compromised safety.

In our previous work, we explored how to improve C-V2X Mode 4 to reduce the AoI and proposed a new performance metric to measure AoI in vehicular networks \cite{10419638}. In this study, we extend our previous work. We focus on a reinforcement learning-based ramp merging scheme, while also considering the impact of C-V2X Mode 4, and propose a new metric to measure the positioning error of control vehicles\footnote{\href{https://github.com/qiongwu86/PPO-Based-Vehicle-Control-for-Ramp-Merging-Scheme-Assisted-by-Enhanced-C-V2X}{The source code has been released at: https://github.com/qiongwu86/PPO-Based-Vehicle-Control-for-Ramp-Merging-Scheme-Assisted-by-Enhanced-C-V2X}}. The main contributions of this paper can be summarized as follows:
\begin{itemize}
	\item[1)] We enhance C-V2X Mode 4 to address its performance limitations by introducing an innovative resource reservation strategy and an improved Short Message Control (SCI) format, which  can resolve communication interruptions and improve transmission reliability.
	
	\item[2)] We formulate a decentralized on-ramp merging scheme with integrated lateral control, thereby eliminating the need for central controllers such as roadside units (RSUs). Our two-step process enables vehicles to adjust their positions and velocities using C-V2X Mode 4 prior to entering the merging zone, after which a reinforcement learning-trained controller guides comprehensive vehicle control.
	
	\item[3)] We devise a novel metric based on the AoI and create a simulation platform using NS3, specifically tailored to evaluate AoI within autonomous driving scenarios where the timeliness of information is crucial for control decisions. This platform integrates a mobility module with vehicular kinematics, allowing for the simultaneous simulation of C-V2X Mode 4 communication and vehicle control.
\end{itemize}

The rest of this paper is organized as follows: Section II reviews related work, Section III details the system model and training objectives for vehicle control, Sections IV and V describe the enhanced C-V2X Mode 4 and our reinforcement learning algorithm, respectively. Section VI presents simulation results and analyses of communication protocols and on-ramp merging performance, later on,  Section VII concludes the study.

\section{Related Work and Motivation}

Recently, the work about improving the performance of C-V2X mode 4 has been investigated in the literature.
In \cite{9528717}, \Athr{Saad} proposed a method based on deep Q network (DQN) algorithm to control the transmission power in physical layer to enhance the performance of C-V2X mode 4.
In \cite{9817813}, \Athr{Gu} proposed a method based on multi-agent reinforcement learning (MARL) to optimize the resource allocation scheme in C-V2X mode 4. In addition, they employed multi-actor-attention-critic (MAAC) to improve the training efficiency and the scalability. 
In \cite{9839138}, \Athr{Ali} proposed to broadcast candidate resources when vehicles are in the procedure of SB-SPS and adjust the resource reselection probability to improve the packet delivery ratio (PDR) in C-V2X mode 4.
In \cite{9569449}, \Athr{Sabeeh} proposed Adaptive Modulation and Collision Detection (AMCD) resource allocation scheme of C-V2X mode 4, where vehicles requires to calculate the Channel Busy Ratio (CBR), and dynamically adjust the Modulation and Coding Scheme (MCS), transmission power, and reselection probability based on the CBR.
In \cite{9759361}, \Athr{Segawa} proposed Interference Prediction and Multi-Interval extension (IPMI), in which an resource allocation scheme adjusts the packet transmission interval based on the position information of surrounding vehicles, to replace SB-SPS in C-V2X mode 4.
In \cite{9323042}, \Athr{Kang} proposed Adaptive Transmission Power and Message Interval Control (ATMOIC) method to adjust the transmission power and interval of C-V2X mode 4 based on the position information of vehicles. 

In the aforementioned works, PDR was used as the performance metric. However, PDR cannot directly reflect the timeliness of information. 
AoI is defined as the difference between the current time and the generation time of the latest received data packet \cite{9380899}. 
It is a metric for assessing information timeliness and is widely used in communication systems which require high information freshness, such as IoV which adoptes C-V2X vehicle communication protocol.
Thus, there has been a few works regarded AoI as optimization objective.
In \cite{parvini2023aoi}, \Athr{Parvini} proposed two algorithms based on multi-agent deep deterministic policy gradient (MADDPG) and used them to train a policy for resource selection in C-V2X. Minimizing average AoI in IoV was used as the objective when training the policy.
In \cite{9839316}, \Athr{Mlika} used NOMA technology at the physical layer to minimize AoI in C-V2X.
However, the above two works optimized the protocol from the perspective of the physical layer without considering the resource allocation scheme in the MAC layer of C-V2X Mode 4. 
In \cite{9214855}, \Athr{peng} proposed a persistent resource allocation scheme in the MAC layer of C-V2X mode 4 called Collision Avoidance based Persistent Schedule (CAPS).
It adds auxiliary information in packet to construct a cooperative scheme when vehicles are allocating resource for transmission in order to reduce the packet collision.


As for ramp merging, there has been many existing works on it and most of them assume a ideal communication condition.
In \cite{xue2022platoon}, \Athr{xue} proposed a platoon-based algorithm for ramp merging control and it can smoothly guide vehicles from the on-ramp to merge into the mainline without significantly affecting the mainline traffic.
In \cite{9565856}, \Athr{Gao} converted the optimal controller that considers lane-changing motivation into a non-linear programming problem in the scenario of ramp merging.
In \cite{10053376}, \Athr{Liu} proposed an on-ramp control architecture for the coexistence of CAVs and human-driven vehicle (HDV). 
The architecture is divided into two layers, where the role of upper layer is to obtain the expected merging point and the lower level obtains the trajectory of the vehicles by solving a QP problem.
With the development of machine learning, more and more work applyed learning based methods to vehicle control in ramp merging scenarios \cite{9016391,kebria2019deep}.
In \cite{9316925}, \Athr{Liu} proposed a lane selection method based on reinforcement learning, which considered the scenario of ramp merging with multiple lanes, and a motion planning algorithm based on time-energy optimal control to guide vehicles movement.
In \cite{9557770}, \Athr{Kherroubi} used artificial neural network (ANN) to predict other vehicles' intentions and used the predicted results as state for reinforcement learning.
However, this scheme is centralized and needs a road side unit (RSU) int the scenario of ramp merging.
In \cite{9986560}, \Athr{Wu} proposed recurrent based twin delayed deep deterministic policy gradient algorithm, which is a kind of RL algorithm combining long short-term memory (LSTM) with the twin delayed deep deterministic (TD3), and used it to control vehicles in ramp merging scenario.
In \cite{9903132}, \Athr{Mahabal} proposed a CAV ramp merging schene that combines Deep Q-Network (DQN) and Deep Deterministic Policy Gradient (DDPG), where the vehicles used DQN to select the lane and DDPG to obtain longitudinal acceleration.
In \cite{9316958}, \Athr{Hu} proposed an ramp merging scheme which combined centralized and decentralized control. 
In this scheme, a RSU placed in the merging area is used for centralized computing of the vehicle's merging order and merging position, and to notify the vehicles accordingly. 
Then, the decentralized controllers, based on the udwadia-kalaba approach and lyapunov stability theory, guided the vehicles to complete the merging process and are employed on each vehicle.
However, the vehicle control scheme in ramp merging conducted in the above works does not specifically include lateral control. 
Thus, in \cite{9729796}, \Athr{Hwang} proposed FSM-RL control scheme by combining finite state machine (FSM) with DRL to control vehicles longitudinally and laterally simultaneously.
In this scheme, the vehicle switches between different states according to the rule of FSM to achieve upper-level control and eventually enters the "Lane-change" mode.
Then control policy based on DRL obtained is used for lower-level control.

All the above works assume that vehicles have communication capabilities to obtain their required information. However, the instability of wireless communication is still worth studying in ramp merging scenarios. Therefore, in \cite{9793718} and \cite{9678122}, communication delay was considered in the control methods for ramp merging. In \cite{9678122}, \Athr{Fang} proposed to model V2I communication delay as a normal distribution through experiments. In addition, in the ramp merging strategy, the RSU first estimates the average delay by communicating with the vehicles multiple times, and then predicts the vehicle's position using this value to obtain more accurate output control. In \cite{9793718}, \Athr{Zhao} also proposed an ramp merging control framework considering communication delays. They conducted comparative experiments under three communication conditions: no time delays, heterogeneous time delays, and homogeneous time delays, and verified their method.

Based on the above discussion, there is currently no work about ramp merging scheme that takes into account the impact of resource allocation scheme used in C-V2X mode 4, and which motivates us to do this work.

\begin{figure}[t]
	\centering
	\includegraphics[trim=0.5cm 0.5cm 0.5cm 0.5cm, clip, width=\columnwidth]{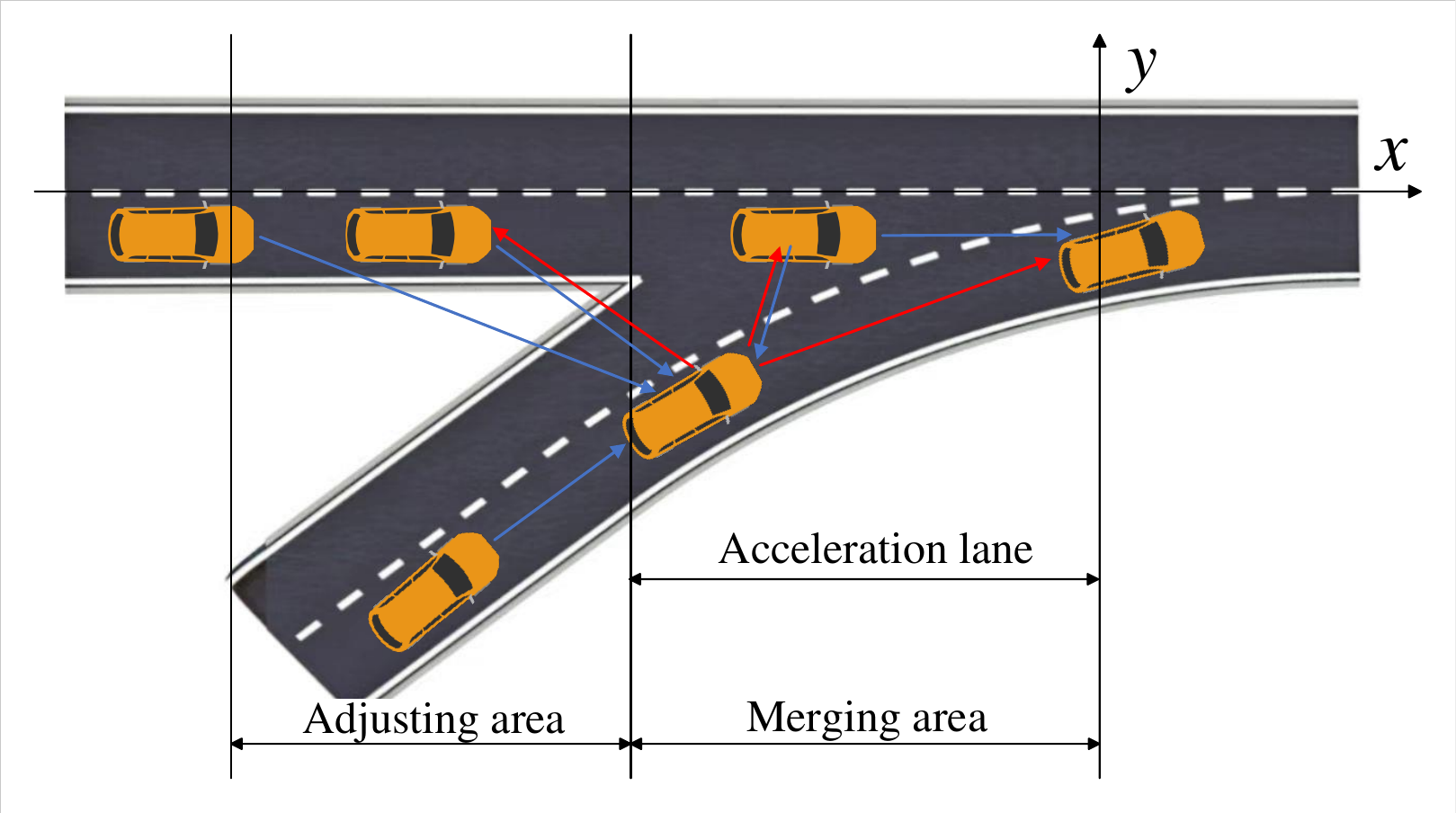}
	\caption{The Schematic Diagram of the Ramp Merging Scenario}
	\label{system-model}
\end{figure}

\section{System Model}

In this section, we first introduce the overall ramp merging scenario. 
Then, we describe the C-V2X Mode 4 resource allocation protocol and the vehicle kinematics model used in this paper. 
Finally, we present the optimization objectives in the last subsection.

\subsection{Ramp Merging Scenario}

As shown in \FigRef{system-model}, we consider the ramp merging scenario and it is divided into two area, i.e., \emph{adjusting area}  and \emph{merging area}.
In \emph{adjusting area}, vehicles need to adjust relative position and velocity to create better condition for the following merging procedure and only the longitudinal control will be taken into account.
As described in \SecRef{intro}, considering vehicles always cannot obtain information about others which is driving on another road, vehicles will adopt C-V2X mode 4 to communicate to each other, where delays and packet losses are present.
Based on the communication provided by C-V2X mode 4, each vehicle will transmit the packet including its real-time information to others and receive the packet sent by others until driving out of \emph{adjusting area}.
After that, the vehicles will drive in \emph{merging area}. 
As for vehicles from main road, they still just need to consider their longitudinal control.
While for vehicles from merging road, they need to consider both longitudinal and lateral control to merge into main road.

\subsection{C-V2X Mode 4 and SB-SPS}

In the physical layer, C-V2X Mode 4 utilizes Long Term Evolution Side Link (LTE-SL) technology to support vehicular communication. The channel is divided into different subframes and subchannels in both time and frequency domains. Each time-domain subframe and frequency-domain subchannel forms a single Subframe Resource (SSR). Each SSR consists of multiple Resource Blocks (RBs). According to LTE-SL, vehicles use two RBs to transmit Side Link Control Information (SCI), and N RBs to transmit Transport Blocks (TBs) containing data, occupying one SSR in total. The value of N is defined in LTE-SL \cite{3gpp.36.331}. Vehicles select SSRs based on the SB-SPS scheme. To ensure communication stability, each vehicle has a Reselection Counter (RC), and it needs to reserve RC SSRs for the transmission of consecutive data packets. 

Assuming that vehicle V needs to transmit data at the subframe \(t_s\), if RC is 0, the vehicle will randomly generate an integer value in the range \([a, b]\) as the RC, and then check whether it can reuse the previous SSR. If reusing fails, the vehicle will select a new SSR and reserve it for the next RC transmissions. 

\begin{figure}[t]
	\centering
	\includegraphics[scale=0.9]{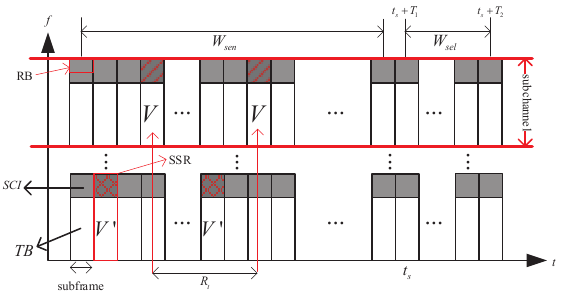}
	\caption{Sidelink of C-V2X Mode 4}
	\label{sidelink:img}
\end{figure}

Assuming the perception window \(W_{sen}\) of the vehicle in the time domain is \([t_s-1000, t_s-1]\), the selection window \(W_{sel}\) is \([ts+T1, ts+T2]\), where \(T1 > 4\) and \(T2\) is in the range \([20, 100]\). Let the set of all SSRs within the selection window be \(S_A\), with a total of \(M_{total}\). Each SSR is represented as \(R_{x, y}\), where \(x\) and \(y\) represent the corresponding subchannel and subframe. The detailed process of SB-SPS is shown in \FigRef{sidelink:img}, and is described as follows.

\begin{itemize}
	\item 1) Vehicle \(V\) obtains the Received Signal Strength Indicator (RSSI) of each SSR from the information received in \(W_{sen}\). It also measures the Reference Signal Received Power (RSRP) based on the corresponding SCI information. These SSRs form the set \(C_{sen}\). The vehicle excludes SSRs that are occupied by other vehicles due to half-duplex operation, as well as SSRs with RSRP exceeding a predefined threshold \(P_{TH}\). These SSRs may experience severe interference from other vehicles. If the number of SSRs remaining in \(S_A\) after exclusion is less than \(0.2 \times M_{total}\), \(P_{TH}\) is increased by 3 dB. The SSRs are then re-filtered based on the new threshold.

	\item 2)  For each remaining SSR in \(S_A\), the vehicle calculates the average RSSI (A-RSSI). The A-RSSI is the average of the RSSIs of all SSRs in \(W_{sen}\) with a 100-subframe interval. This measures the performance of the corresponding sub-channel over different time subframes.

	\item 3) The vehicle arranges the SSRs in \(S_A\) in ascending order based on A-RSSI. It then moves the SSRs into \(S_B\) sequentially. Once the number of SSRs in \(S_B\) exceeds \(0.2 \times M_{total}\), the vehicle randomly selects one SSR from \(S_B\) to transmit data. The subsequent RC SSRs are retained for further RC transmissions until the RC counter reaches zero, triggering a re-selection or the continued use of SSRs in \(S_B\).
\end{itemize}

\subsection{Kinematic Bicycle Mode}

The mobility model of vehicles is kinematic bicycle model in \cite{polack2017kinematic}, which is described as
\begin{equation}
\left\{
\begin{array}{*{2}{cl}}
x_{t+1}&=x_{t}+v_{t}cos(\Phi_{t})\Delta T\\
y_{t+1}&=y_{t}+v_{t}sin(\Phi_{t})\Delta T\\
v_{t+1}&=v_{t}+a\Delta T\\
\Phi_{t+1}&=\Phi_{t}+{v_{t} \delta_t \Delta T}/{L}
\end{array}
\right.,
\end{equation}
where $x$ and $y$ is the coordinate of vehicle, $\Phi$ and $v$ are the heading angle and velocity, respectively.
The inputs of this model is acceleration and steering, which are represented as $a$ and $\delta$, respectively.
In addition, $\Delta T$ is the sample step and $L$ is wheel base.
The vehicle is equivalent to a rectangle with a length of $L$ and a width of $W$.

\subsection{Optimization Objectives of Vehicle Control in Ramp Merging Areas}

In merging process, the controller minimizes fuel consumption and maximizes comfort while ensuring security.
Minimizing fuel consumption is achieved by optimizing the square of acceleration, and a lower of derivative of acceleration and heading angle with respect to time means a higher degree of passengers confort\cite{NTOUSAKIS2016464,rios2015online}.
Therefore, the objective is
\begin{equation}\label{opt-problem}
\begin{aligned}
	\min_{a_i^j, \delta_i^j} & \sum_{i \in [1,N]} \sum_{j \in T_{i}} \left[ a_i^j \right]^2 + \frac{a_i^{j+1} - a_i^{j}}{\Delta T} + \frac{\Phi_i^{j+1} - \Phi_i^{j}}{\Delta T}, \\
	\text{s.t.} \\
	& \quad \text{(policy-C)} \\
	& \quad \left\{
	\begin{aligned}
		v_{min} & \leq v_{i}^{j} \leq v_{max}, \\
		a_{min} & \leq a_{i}^{j} \leq a_{max}, \\
		\delta_{min} & \leq \delta_{i}^{j} \leq \delta_{max}, \\
	\end{aligned}
	\right. \\
	& \quad \text{(communication-C)} \\
	& \quad \left\{
	\begin{aligned}
		&C_{v_i} \cap C_{v_i^{'}} = \emptyset \quad \text{or} \quad RSPP_{v_{i}^{'}} < P_{TH}, \\
		&AveRSSI(R_i) \geq S_{AveRSSI}(0.2*M_{total}),
	\end{aligned}
	\right. \\
	& \quad \text{(soft-C)} \\
	& \quad \left\{
	\begin{aligned}
		d_{i,k}^j & \leq d_{i,k}^{safe}, \\
		d_{i,E}^j & \leq d_{i,E}^{safe}, \\
	\end{aligned}
	\right. \\
	& \quad \text{(hard-C)} \\
	& \quad \left\{
	\begin{aligned}
		&\sum_{i \in [1, N]} \sum_{\substack{\mu \in [1, N] \\ \mu \neq i}} \mathbb{I}(B_{v_i} \cap B_{v_\mu} \neq \emptyset) = 0, \\
		&\sum_{i \in [1, N]} \mathbb{I}(B_{v_i} \cap B_{r} \neq \emptyset) = 0,
	\end{aligned}
	\right.
\end{aligned}
\end{equation}
where \textit{police-C} represents the constraints on the vehicle control strategy. The variables \( a_i^j \) and \( \delta_i^j \) are the optimization variables, representing the acceleration and steering angle of vehicle \( i \) at time \( j \), respectively. The limits on velocity, acceleration, and steering angle are denoted as \( v_{min} \), \( v_{max} \), \( a_{min} \), \( a_{max} \), \( \delta_{min} \), and \( \delta_{max} \).
Our improvement on the C-V2X model 4 protocol is trying to ensure communication quality, with corresponding communication constraints expressed as \textit{communication-C}. Here, \( C_{v_i} \) represents the communication resources reserved by vehicle \( v_i \), \( v_i' \) represents the vehicle communicating with \( v_i \), and \( RSPP_{v_i'} \) represents the reference signal received power of vehicle \( v_i' \). 
Therefore, the first equation ensures that two communicating vehicles cannot reserve the same resources or satisfy the condition that the reference signal received power exceeds the threshold \( P_{TH} \). \( AveRSSI \) denotes the average received signal strength of the channel, and the second equation states that the average received signal strength of the resources selected by vehicle \( i \) should rank in the top 0.2 * \( M_{total} \) of all available resources. \( S_{AveRSSI} \) represents the set of all available resources sorted in descending order of RSSI.

The safety constraints can be divided into two parts: \textit{soft constraints (Soft-C)} and \textit{hard constraints (Hard-C)}. Soft-C require each vehicle to maintain a safe distance from other vehicles or the road edges during driving to avoid collisions or interference. At time \( j \), the distance between vehicle \( i \) and vehicle \( k \) is denoted as \( d_{i,k}^j \), while the distance from the rectangle representing vehicle \( i \) to the nearest point of the road edge is denoted as \( d_{i,E}^j \). The minimum safe distance between vehicle \( i \) and vehicle \( k \) is represented as \( d_{i,k}^{safe} \), and the minimum safe distance between vehicle \( i \) and the road edge is \( d_{i,E}^{safe} \).
Hard-C require that the rectangle of each vehicle must not collide with other vehicles or the road edge. The collision between vehicles and the road is easily detectable, and the Separating Axis Theorem (SAT) is used to detect collisions between vehicles \cite{huynh2009separating}. \( B_{v_i} \) represents the spatial range of vehicle \( i \), and \( B_r \) represents the spatial range outside the road. If the intersection between \( B_{v_i} \) and \( B_{v_\mu} \) is empty, it indicates that vehicle \( v_i \) and vehicle \( v_\mu \) will collide. If the intersection between \( B_{v_i} \) and \( B_r \) is empty, it indicates that the vehicle does not collide with the road. $\mathbb{I}$ is an indicator function that takes a value of 1 if the condition inside is satisfied, and 0 otherwise.

\section{Enhanced C-V2X mode 4}

In this section, we introduce an improved method for C-V2X Mode 4, aiming to enhance communication reliability and optimize the AoI. We first propose an Enhanced SB-SPS (ESB-SPS) algorithm, which addresses the issue where vehicles may select SSRs that are in the same time subframe as the target vehicle. This can lead to long communication failures due to half-duplex operation and reserved resources. The specific process is shown in \AlgRef{pseudocode of ESB-SPS}. Similar to SB-SPS, vehicle $V$ needs to decide the selected SSR based on \( W_{sen} \), \( C_{sen} \), \( S_A \), RC, and $P_{TH}$, with no less than $0.2 \times M_{total}$ SSRs retained in the $S_B$.

Next, we provide a detailed explanation of the ESB-SPS method. First, the vehicle needs to exclude inappropriate SSRs from $S_A$. Due to the half-duplex mechanism, when two vehicles select SSRs located in the same subframe but on different subchannels, they cannot communicate with each other. In this case, due to the characteristics of semi-persistent scheduling, vehicles operating under standard C-V2X Mode 4 will continue to attempt these failed transmissions until the RC (resource count) of one of the vehicles drops to 0 and the SSR is re-selected, resulting in long communication failures that impact information freshness. To address this issue, we propose a new resource reservation method that ensures no overlap between the SSR reserved by the vehicle and the SSR reserved by the communication target vehicle, thereby avoiding consecutive communication failures. When vehicle $V$ excludes an SSR from $S_A$, it first calculates which SSRs will be reserved if this SSR is selected, and stores them in the set $C_V$. The reserved SSRs can be computed using the mapping function $CO^{[i]}$:

\begin{equation}
\begin{aligned}
	&CO^{[i]}[(x,y,z)]=\\
	&((x+i*\frac{R_t}{10})mod(1024), (y+i*z)mod(10), z),\\
\end{aligned}
\end{equation}
where \( x \), \( y \), and \( z \) represent the frame number, subframe number, and subchannel number of the current SSR (denoted as \( R \)), respectively. Each System Frame Number (SFN) cycle consists of 1024 frames, and each frame contains 10 subframes (i.e., 10 ms). Assume that the environment contains \( SC \) subchannels. Thus, \( x \in [0, 1023] \), \( y \in [0, 9] \), and \( z \in [0, SC-1] \).

\begin{algorithm}
	\setstretch{0.65}
	\LinesNumbered
	\caption{ESB-SPS Pseudocode}
	\label{pseudocode of ESB-SPS}
	\small
	\KwIn{
		$W_{sen}$, $C_{Sen}$, $S_{A}$, RC, $P_{TH}$
	}
	\KwOut{SSR $R_{new}$}
	$M_{total} = |S_{A}|$;\\
	\While{$S_{A} > 0.2 \times M_{total}$}
	{
		\ForEach{$R \in S_{A}$}
		{
			$C_{V} = \emptyset$;\\
			\For{$i \in [0, RC-1]$}
			{
				append $CO^{[i]}(R) \KwTo C_{V}$;
			}
			\ForEach{$R' \in C_{Sen}$}
			{
				Get $RC'$ and $RSRP'$ of $R'$;\\
				$C_{V'} = \emptyset$;\\
				\For{$i \in [0, RC'-1]$}
				{
					append $CO_{R'}^{[i]} \KwTo C_{V'}$;
				}
				\If{$C_{V} \cap C_{V'} \neq \emptyset$ and $RSRP' > P_{TH}$}
				{
					remove $R$ from $S_{A}$;\\
					\KwGoto {$18$}
				}
			}
		}
		$P_{TH} \gets P_{TH} + 3dB$;\\
	}
	Calculate $A-RSSI$ for each SSR in $S_{A}$;\\
	\While{$S_{B} < 0.2 \times M_{total}$}
	{
		Move $R$ with the smallest $A-RSSI$ to $S_{B}$;
	}
	$R_{new} \gets$ randomly select from $S_{B}$;\\
	return $R_{new}$;
	\label{Over}
\end{algorithm}

After collecting the SSRs that are reserved for the current SSR \( R \), we need to obtain the RC and RSRP values corresponding to each SSR \( R' \) in the set \( C_{sen} \). In the standard C-V2X Mode 4, the SCI of \( R' \) includes the RSRP, but vehicle \( V \) cannot directly obtain the RC of \( R' \). Instead, it estimates the RC using the predicted value \( \lceil 100/R_t \rceil \). This results in \( V \) being unable to accurately determine whether the SSR has been reserved by other vehicles, potentially selecting an inappropriate SSR and increasing the probability of transmission failure.

To accurately determine if an SSR is reserved by other vehicles, we aim to obtain the RC value of other vehicles. In standard C-V2X, the SCI includes the vehicle's priority and retransmission information, occupying 8 bits. According to the protocol specifications, when \( R_t = 20 \), the RC range is \( [25, 75] \). In this case, transmitting the RC requires at least \( \left\lceil \log_{2}{75-25+1} \right\rceil = 6 \) bits. 
Considering that retransmission information increases the load on the vehicular network\footnote{Vehicles using retransmission mechanisms transmit a data packet twice \cite{3gpp.36.213}.}, a higher number of vehicles may lead to network congestion, reducing communication reliability. Therefore, we assume that vehicles have the same priority and do not use retransmission mechanisms. We allocate the 8 bits used for retransmission and priority to store the RC value. In this case, vehicles can directly obtain the RC value from the SCI, accurately excluding SSRs reserved in \( S_A \). The proposed SCI format is shown in \TableRef{sci}.

\begin{table}[htbp]
	\centering
	\caption{Proposed SCI format}
	\label{sci}
	\resizebox{0.45\textwidth}{!}{
	\begin{tabular}{ccc}
		\toprule
		Index & Item & bits \\
		\midrule
		1 & Resource Reservation &  4 \\
		2 & Frequency Resource Location & $\log(SC(SC+1)/2)$ \\
		3 & MCS & 5 \\
		4 & Transmission Format & 1\\
		5 & Reserved & $14 - \log(SC(SC+1)/2)$ \\
		6 & RC & 8  \\
		\bottomrule
	\end{tabular}}
\end{table}

The vehicle $V$ obtains $RC'$ from $R'$ and calculates the SSR reserved by vehicle $V'$ using the $CO$ function. The reserved SSR is stored in the set $C_{V'}$. If $C_{V} \cap C_{V'} \neq \emptyset$, it indicates that selecting the current SSR $R$ will result in an overlap with another vehicle's reserved SSR. If $RSRP' > P_{TH}$, it indicates significant interference between vehicle $V$ and vehicle $V'$, and thus the current SSR must be excluded. After processing the current SSR, the algorithm proceeds to check the next SSR in $S_{A}$ to see if it will be excluded. Once all SSRs have been checked, if the remaining SSRs are less than $0.2 \times M_{total}$, $P_{TH}$ increases by 3dB, and the above process repeats. Otherwise, the vehicle calculates the $A-RSSI$ for each SSR, sorts them, and moves them to $S_B$. The vehicle then randomly selects an SSR from $S_B$ as $R_{new}$. The $CO$ function calculates the SSR that will be reserved based on the $RC$ value.

When calculating the A-RSSI, vehicles can directly select the SSRs from \( W_{sen} \) that can be mapped to the current SSR \( R \) through the CO function. Assuming \( R \) is located at \( (x, y, z) \), the function for calculating the A-RSSI can be defined as follows:
\begin{equation}\label{asl-rssi}
	\begin{aligned}
    AveRSSI(R)= \frac{1}{|C_{RSSI}(R)|}\sum_{r \in C_{RSSI}(R)}^{}[RSSI(r)],
	\end{aligned}
\end{equation}
where
\begin{equation}
	   \begin{aligned}
		&C_{RSSI}(R)=\\
		&\{(a,b,c) \in W_{sen}|CO^{[i]}[(a,b,c)]=(x,y,z),\exists i \in N\}.
	   \end{aligned}
\end{equation}

$|C_{RSSI}(R)|$ is the number of elements in set $C_{RSSI}(R)$ and $RSSI(r)$ is RSSI of SSR $r$.


\section{Vehicle Control Method}\label{V.Vehicle Control Method}
The vehicle controller is divided into two parts, the first part is based on Cooperative Adaptive Cruise Control (CACC) and its own longitudinal control, while the second one is based on reinforcement learning with both lateral and longitudinal control. 
As for the vehicles from main road, they will use the first controller to control the longitudinal position and velocity in both adjusting and merging area.
However, the vehicles from merging road will use the first controller when driving in adjusting area and switch into the second controller after driving into merging area because they need to merge into main road.

\subsection{CACC Control}\label{1.CACC control}
CACC is a popular model applied to the study of autonomous vehicle following. The CACC model we use comes from SUMO, which consists of four mode and they are activated at different situations. The four modes are:

\begin{itemize}
\item speed control mode
\item gap-closing control mode
\item gap control mode
\item collision avoidance control mode
\end{itemize}

Speed control mode is activated to maintain the vehicle at a pre-set speed while acceleration is calculated as 
\begin{equation}
a_{t}=k_{1}*(v_{d}-v_{t}),
\end{equation}
where $k_1$ is the parameter and $v_{d}$ is the desired velocity.
As for the other three modes, the desired velocity at next time step is calculated as 
\begin{equation}
	v_{t+1}^d=v_{t}+k_{2}^i * P_{err} + k_{3}^i * V_{err},
\end{equation}
where $k_2^i$ and $k_3^i$ are parameters in mode $i$, where $i=2,3,4$ corresponding to the last three mode.
$P_{err}$ at time $t$ is calculated as $(x_{t,p} - x_{t}) - T_{h} * v_{t}$, where $x_{t,p}$ and $x_t$ are previous vehicle and ego longitudinal position, respectively, and $T_{h}$ is headway time. Similarly, $V_{err}$ at time $t$ can be calculated as \((v_{t, p} - v_t) - T_h * a_t\).
After getting the described velocity, the vehicle should calculate the next acceleration as $\frac{v_{t+1}^d - v_t}{\Delta T}$ and limit the acceleration in the bound $[a_{min}, a_{max}]$.
Then the acceleration is put into the discrete longitudinal dynamic, i.e., $v_{t+1} = v_{t} + a * \Delta T$, and the velocity $v_{t+1}$ will also be limited in $[v_{min}, v_{max}]$.

Based on the description above, CACC needs the previous vehicle's information, including position and velocity, to calculate $P_{err}$ and $V_{err}$ and then calculate the acceleration.
To get the information of previous vehicles, each vehicle will use V2X communication technology to communicate to each other and find the vehicle closest to it in the longitudinal direction.
However, the information is not instantaneous, i.e., AoI is not $0$ even without transmission failures.
Specifically, AoI comes from three sources:
\begin{itemize}
	\item There is a lag from the MAC layer to the application layer(in NS3, the length is 4ms).
	\item The data packet was not successfully received, and the age of the information at this time is related to the interval between sending packets and the number of consecutive unreceived packets. If $n$ consecutive packets are not received, the delivery interval is $Rsvp$. Then the AoI will increase by $Rsvp*n$ ms.
	\item The time difference between the control time and the packet receiving time. Regardless of the first two sources of information age, the frequency of contracting is $Rsvp$. When the value is uniformly distributed within $[1,Rsvp]$ between.
\end{itemize}

Therefore, there is an error in position when the vehicle control itself.  
In order to reduce this error, we use AoI to predict the real position of other vehicles.

To calculate AoI, the vehicle adds the packet generation time when creating the packet including the information of itself. 
Specifically, the format of the packet is $\{ID, x, y, v, \theta, ROAD, TS\}$.
$ID$ is the vehicle ID number of the sender, which is used to identify the source of the data packet. 
$x$ and $y$ are the coordinates of the rear wheel of the bicycle model, $v$ is the speed of the vehicle, $\theta$ is the vehicle body angle, $TS$ is the timestamp of the packet, and $ROAD$ is the section of road the vehicle is traveling on. 
$ROAD$ can take one of three values, namely $MAIN$, $MERGE$, and $MERGING$, which respectively represent the main road, the merging road, or the already passed merging point.
Each vehicle has a dedicated buffer for storing received data packets from each sender.
Whenever a new data packet is received, it will replace the last one from the sender of the packet and store in the buffer.
Assuming that vehicle $A$ reaches the control moment $t_{c}$, it first estimates the current position of each vehicle in the scene, and then projects using the estimated position. 
Assuming that there is another vehicle $B$ in the scene, vehicle $A$ obtains the projection of $B$ by the following two steps:

\begin{itemize}
	\item Position correction
	
	The format of packet in vehicle A's buffer about vehicle B at this time is 
	\begin{equation}
	P_A^B=[B, x_{B},y_{B},v_{B},\theta_{B}, ROAD_{B}, TS_{B}].
	\end{equation}
	A first calculates the information age about B, $AoI_{A,B}=t-TS_{B}$. Assuming that B maintained a constant speed between the last packet generation and $t_{C}$, the distance traveled by the vehicle is $AoI_{A,B}*v_{B}$ meters. Assuming the corrected coordinates are $(x_c,y_c)$.
	\item Projection
	
	Since there is no lateral control in CACC, we project the vehicle's position and velocity onto a one-dimensional space. For any vehicle, the projected coordinates are equal to the distance traveled along the current road to the $y$-axis, as shown in \FigRef{system-model}.

\end{itemize}

After vehicle $A$ calculated the projections of all the vehicles based on the information in the buffer, it sorts them based on their distance from A. 
If there is no vehicle in front $A$, it will use the first mode in CACC, i.e., speed control mode.
Else, it will find the closest vehicle $R$ in front of it and uses the information about $R$ to calculate $P_{err}$ and $V_{err}$.
And then control itself in the following 3 mode in CACC.

\subsection{Reinforcement Learning Control Method}\label{2.RL}

Proximal policy optimization (PPO), a reinforcement learning algorithm, can solve continuous control problems with continuous action spaces\cite{schulman2017proximal}. 
It uses important sampling techniques to implement off-policy reinforcement learning, which improves the utilization of data. 
At the same time, this algorithm does not require the use of additional target networks like the DDPG algorithm and can output continuous actions. 
Therefore, we use the PPO algorithm to solve the problem. 
After passing the adjusting area, the vehicles from the merging road will use a controller trained by PPO to control themselves. 
In this section, we first introduce the model, including action space, state space, and rewards, etc. 
Then we introduce the training process of PPO.

\begin{itemize}

\item State Space

\quad
In our scenario, the state $S$ of each agent is divided into three parts: $S_{ego}$, $S_{prev}$, and $S_{foll}$. $S_{ego}$ is used to describe the state of the agent itself, which is defined as
	\begin{equation}
		S_{ego}=[x, y_{r}, y_{f}, v, \Phi],
	\end{equation}
where $x$ represents the x coordinate. 
To better describe the $y$-axis state of the vehicle, we include the $y$-axis coordinates both of the rear and front wheels of the bicycle model in its own state, denoted as $y_{r}$ and $y_{f}$ respectively. 
Here, $y_{r}$ is equal to the $y$ coordinate of the bicycle model, while $y_{f}=\sin(\Phi)*wheelbase+y_{r}$. 
$v$ and $\Phi$ represent the speed and body angle of the bicycle model, respectively.
$S_{prev}$ is used to describe the state information of the nearest vehicle located in front of the vehicle. This is defined as $[\Delta_{prev}^{x},\Delta_{prev}^{v}]$. 
The first term represents the difference in x-axis between the previous vehicle and the ego vehicle. 
To describe the relative speed between the lead and following vehicles on the $x$-axis, the second term is defined as $\Delta_{prev}^{v}=v*\cos(\Phi)-v_{prev}*\cos(\Phi_{prev})$. 
Here, $v_{prev}$ and ${\Phi_{prev}}$ respectively represent the speed and body angle of the lead vehicle. 
$S_{foll}$ is similar to $S_{prev}$, used to describe the state of the nearest vehicle located behind the vehicle and it is represented as $[\Delta_{foll}^{x},\Delta_{foll}^{v}]$. 

\item Action Space

\quad
We consider both longitudinal and lateral control, therefore the state space is defined as $[a,\delta]$, representing the acceleration and steering angle input to the bicycle model, respectively. 
The range of $a$ is $[a_{min},a_{max}]m/s^2$. The range of $\delta$ is $[-15^{\circ}, +15^{\circ}]$.

\item Reward Function

\quad 
The reward function is related to the objective in \EquRef{opt-problem}.
Firstly, considering \emph{Hard-C} is related to safety and should not violate anytime.
Therefore, if the vehicle violates the \emph{Hard-C}, the episode of it terminates and the reward is calculated as
\begin{equation}\label{fail-termination-reward}
	r_1=-C_T^1-k_1^r |x| - k_2^r (|y_r| + |y_f|),
\end{equation}
where $C_T$ is a larger constant and express a severe punishment. $|x|$ is the absolute of x-axis coordinate at the terminate time. 
The smaller the $|x|$ value at the terminate time, the closer vehicle is to the endpoint, resulting in a smaller penalty.  
This can better tell the training direction of the algorithm and improve convergence speed.
The last part is similar, it express the coordinate of y-axis of vehicle at terminate time and it can guide the vehicle to try to drive belong the center of the road.
$k_1^r$ and $k_2^r$ are two parameters.

\quad 
If the vehicle successfully passes without collision after performing an action, the episode for that vehicle ends, and the reward at this point is 
\begin{equation}\label{success-termination-reward}
	r_2=C_T^2 - k_3^r|y_r| - k_4^r|\theta| + k_5^r \sum a_{t} + k_6^r \sum \delta_{t},
\end{equation}
where $C_T^2$ is a larger positive value to express the succeed for merging.
In addition, to encourage the vehicle to maintain a smaller body angle and drive in the center of the road as much as possible when driving out of merging area, $|y_r|$ and $|\theta|$ are added.
Moreover, to minimize the objective in \EquRef{opt-problem}, $\sum a_{t}$ and $\sum \delta_{t}$ are added to minimum the fuel consumption.
While, $k_3^r$, $k_4^r$ and $k_5^r$ are still parameters to control the proportion of different items.

\quad 
In the usual case, the vehicle neither collides nor successfully passes through. 
The reward at this time is $r_3=r_{ego}+r_{other}$, where $r_{ego}$ is related to the second constraint in \EquRef{opt-problem} and can be expressed as $r_{ego}=k_{ego}^x r_{x} + k_{ego}^y r_{y} + k_{ego}^\theta r_{\theta} + k_{ego}^{act} r_{act}$, and 
\begin{equation}
\left\{
\begin{array}{*{2}{cl}}
r_{x}&=-|\frac{x}{L_{a}}|\\
r_{y}&=(1-abs(x/L_a))*(R_Y(y_r)+R_Y(y_f))\\
r_{\theta}&=k_{\theta}^1*(\theta)^2+k_{\theta}^2*abs(\theta - \theta^{'})\\
r_{act}&=F_a(a,a_{min},a_{max}) + F_a(\delta,\delta_{min},\delta_{max})\\
\end{array}
\right..
\end{equation}
$r_x$ is related to the vehicle's longitudinal position and it encourage the vehicle to drive as far as possible without collision.
In $r_{y}$, 
\begin{equation}
	R_Y(y)=
	\begin{cases}
	abs(\frac{y}{1.5R_{w}}), y < 0 \\
	abs(\frac{y}{0.5R_{w}}), else
	\end{cases},
\end{equation}
and this value guide the vehicle drive along the center of the main road, where $R_w$ is the width of road.
We add $(1-abs(x/L))$ in $r_{y}$ because we want the vehicle to stay closer to the center of the road as it approaches the end point. 
This term makes $r_{y}$ close to 0 when the vehicle just enters the acceleration section. 
As the vehicle approaches the end point, $(1-abs(x/L))$ becomes larger. 
Therefore, the vehicle will try to drive closer to the center of the road at this point. 
Without this term, the vehicle may output a large steering angle to reduce $(R_Y(y_r)+R_Y(y_f))$ as soon as possible, which will ultimately lead to too fast a turning speed and affect ride comfort. 
Additionally, in our scenario, the range of $y$ values is $[-1.5*R_{w},0.5*R_{w}]$. 
To prevent uneven positive and negative values of $y$ from offsetting the learning target, we define $R_Y$ as a segmented function.
Then, $\theta^{'}$ is the vehicle's body angle from the previous time step. 
We want the vehicle to travel as parallel to the lane as possible, so we add the first term. 
Additionally, to minimize the shaking of the vehicle and increase comfort as much as possible, we add the second term. 
$r_{act}$ is related to the input. 
As there are two parts to the action, acceleration $a$ and steering angle $\delta$, this term is defined as the sum of two parts and $F_a$ is defined as follows:
\begin{equation}
	F_a(x,MIN,MAX)=
	\begin{cases}
		(\frac{x}{x_{MIN}}),x \le 0 \\
		(\frac{x}{x_{MAX}}),x > 0
	\end{cases}.
\end{equation}
This term is used to make the output of the policy as small as possible. 
A smaller term can not only make the vehicle's speed and body angle more stable, but also minimize energy consumption as much as possible.

\quad
$r_{other}$ is related to other nearby vehicles. We take into account the relative position and velocity of two closest cars in front and behind this vehicle.
Specifically, $r_{other}=F_o^p + F_o^f$, where 
\begin{equation}\label{other-reward}
	F_o^p=
	\begin{aligned}{}
	\overbrace{
	k_o^{p,1} \times {
	\left\{
	\begin{array}{*{2}{cl}}
	max(-(d_p/k_{d}^p)^2,-1), d_p < 0\\
	e^{-d_p}-1, d_p \geq 0
	\end{array}
	\right.
	}}^{position} \\
	+ 
	\overbrace{k_o^{p,2} e^{-|v_{ego}cos(\Phi_{ego})-v_{p}cos(\Phi_{p})|-1.0}}^{velocity},
	\end{aligned}
\end{equation}
where $v_{ego}$ and $v_{p}$ are the velocity of ego vehicle and previous vehicle, and $\Phi_{ego}$ and $\Phi_{p}$ are the body angle of ego and previous vehicle, respectively.
In \EquRef{other-reward}, the first item is the reward related to relative position between ego to previous vehicle. 
$d_p=(x_{ego}-x_{p}) - T_h v_{ego}sin(\Phi_{ego})$, it is the error between read distance to expected distance calculated by headway time $T_h$ and longitudinal speed.
If $d_p < 0$, it means that the safe distance is not meet, thus the penalty should be more severe.
We use a quadratic function to represents it and limit it to -1 to prevent excessive rewards. $k_d^p$ is the parameter to control the value of $d_p$ when the function reaches its minimum.
When $d_p \geq 0$, it means that the safe distance is meet but a little big. 
If $d_p$ is too large, it means a much longer longitudinal distance between two vehicles.
At this time, we want ego vehicle to accelerate to approach previous vehicle, we give a smaller penalty, i.e., $e^{-d_p}-1$, which will will tend towards -1 as $d_p$ approaches infinity.
The second in \EquRef{other-reward} is related to relative velocity of ego vehicle to previous vehicle.
$v_{ego}cos(\Phi_{ego})$ is the longitudinal velocity of ego, and we want ego vehicle to try to keep the same longitudinal velocity with previous, i.e., $v_{p}cos(\Phi_{p})$.
$F_o^f$ is similar with $F_o^p$ and it is calculated as \EquRef{other-reward} where replace $p$ to $f$.
\end{itemize}

\subsubsection*{PPO algorithm}\label{2.RL control}

In reinforcement learning, the trajectory of a single movement of an agent can be described as $\tau=\{s_1,a_1,s_2,a_2,\dots\}$. Assuming the policy function is $\pi_{\theta}(a|s)$, where $\theta$ is a parameter. It represents the probability of performing action $a$ in state $s$. The goal of reinforcement learning is to maximize 
\begin{equation}
	J(\theta)=E_{\tau \sim \pi_{\theta}} [R(\tau)],
\end{equation}
where $R(\tau) = \sum_{t=1}^{T} \gamma^{t} r(s_t,a_t)$, it represents the return of $\tau$. $\gamma \in (0,1)$ is used to prevent $R(\tau)$ from being unbounded when $\tau$ is infinitely long. $r(s_t,a_t)$ is the reward obtained after performing action $a_t$ in state $s_t$. $J(\theta)$ represents the expected return under the parameter $\theta$. By using the EGLP lemma and adding a baseline, the gradient of the objective function can be approximately solved with the following formula,
\begin{equation}
	\nabla_{\theta} J(\theta) \approx E_{{(s_t,a_t) \sim \pi_{\theta}}}[\nabla_{\theta} \log \pi_{\theta} (a_t|s_t) (R(s_t,a_t) - V_{\omega}(s_t))] \label{grad01},
\end{equation}
where $R(s_t,a_t)=\sum_{i=0} \gamma^{i} r(s_{t+i},a_{t+i})$. We can update the parameter $\theta$ using $\theta^{'} \gets \theta + \alpha \nabla_{\theta} J(\theta)$. $V_{\omega}(s_t)$ is the value function. This function also uses a neural network to approximate, where $\omega$ is the parameter of the value function. We optimize the value function by optimizing its loss function $Loss_V(\omega)$. $Loss_V(\omega)$ is defined as 
\begin{equation}
		Loss_V(\omega) =  E_{(s_t,a_t) \sim \pi_{\theta}}[(R(s_t,a_t)-V_{\omega}(s_{t}))^{2}]. \label{obj01} \\
\end{equation}

In \EquRef{grad01}, the gradient is an expected value, which can be calculated by sampling. However, $\tau \sim \pi_{\theta}$ means that the data needs to be sampled using $\pi_{\theta}$. After one parameter update of $\theta$, the data sampled using $\pi_{\theta}$ is no longer available, which reduces the efficiency of data utilization. To solve this problem, the PPO algorithm uses the method of important sampling. In the PPO algorithm, the gradient is calculated using the following formula

\begin{equation}
	\nabla_{\theta} J(\theta) = E_{(s_t,a_t) \sim \pi_{\theta_{old}}}[\frac{\pi_{\theta} (a_t|s_t)}{\pi_{\theta_{old}} (a_t|s_t)} Adv \nabla_{\theta} \log p_{\theta}(a_t|s_t)]. \label{grad02}
\end{equation}

Where $Adv$ is the advantage function. When this function is greater than 0, it will increase the probability of $\pi_{\theta}$ taking action $a_{t}$ in state $s_{t}$, and vice versa. This function is defined as

\begin{equation}
	Adv = [R(s_{t},a_{t}) - V_{\omega}(s_t)]. 
\end{equation}

In \EquRef{grad02}, $(a_t,s_t) \sim \pi_{\theta_{old}}$ indicates that when using sampling to approximate $\nabla_{\theta} J(\theta)$, $\pi_{\theta_{old}}$ can be used for sampling. In other words, data can be used multiple times to improve sampling efficiency. In addition, if $\pi_{\theta_{old}}$ is too small, it will lead to a large gradient, causing unstable learning. Therefore, before calculating $\nabla_{\theta} J(\theta)$, the target function $J(\theta)$ is clipped to solve this problem. The final objective function is obtained as

\begin{equation}
	\begin{aligned}
	J_{PPO}(\theta) = &E_{(s_t,a_t) \sim \pi_{\theta_{old}}}\{\min [\frac{\pi_{\theta} (a_t|s_t)}{\pi_{\theta_{old}} (a_t|s_t)}Adv,\\
	& CLIP(\frac{\pi_{\theta} (a_t|s_t)}{\pi_{\theta_{old}} (a_t|s_t)}, 1-\epsilon, 1+\epsilon)Adv] \} \label{obj02}
	\end{aligned}.
\end{equation}

\begin{figure*}[htbp]
	\centering
	\includegraphics[trim=0.5cm 0.5cm 0.5cm 0.5cm, clip, width=\textwidth]{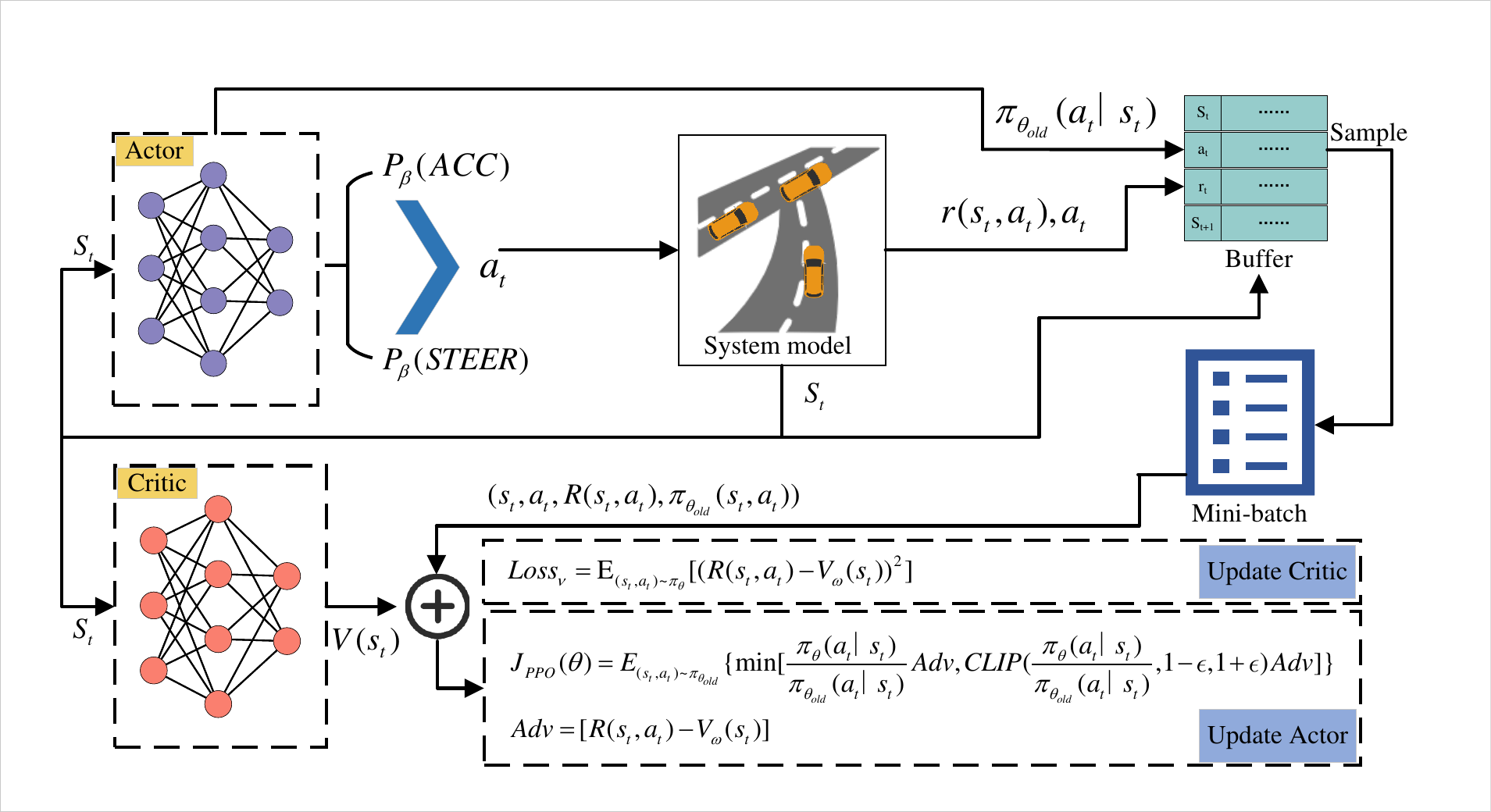}
	\caption{Schematic Diagram of Proximal Policy Optimization}
	\label{ppo}
\end{figure*}

Fig.~\ref{ppo} illustrates the training process of the PPO algorithm. Below, we introduce the specific steps of our training and testing.

\begin{algorithm}[]
    \setstretch{0.75}
    \LinesNumbered
    \caption{Training Stage for the PPO based Framework}
    \label{alg2}
    \KwIn{$\theta$, $\omega$}
    \KwOut{optimized:$\theta^{*}$,$\omega^{*}$}
	Randomly initialize the $\theta$, $\omega$;\\
	Initialize replay experience buffer $\mathcal{R}$;\\ 
	\For{epoch from 1 to $epoch_{max}$}
	{
		clear buffer $\mathcal{R}$;\\
		\Do{buffer is not full}
		{
			Initialize environment, add vehicles to group $\mathcal{G}$;\\
			$t=0$;\\
			\Do{$\mathcal{G} \ne \emptyset$}
			{
				$t=t+1$;\\
				Initialize $\mathcal{H} = \emptyset$;\\
				\For{$v_{i}$ \Kwin $\mathcal{G}$}
				{
					\eIf{$v_{i}$ in CACC mode}
					{
						Generate $a_{i,t}$ by CACC;\\
					}
					{
						Sample $a_{i,t}$ from $\pi_{\theta}(a|s_{i,t})$;\\	
						Append $v_{i}$ to $\mathcal{H}$;\\
					}
				}
				\For{$v_{i}$ \Kwin $\mathcal{G}$}
				{
					Execute $a_{i,t}$;\\
				}
				\For{$v_{i}$ \Kwin $\mathcal{G}$}
				{
					\If{$v_{i}$ \Kwin $\mathcal{H}$}
					{
						Observe $r(s_{i,t},a_{i,t})$ and $s_{i,t}^{'}$;\\
						Save $(s_{i,t},a_{i,t},r(s_{i,t},a_{i,t}),s_{i,t}^{'},p_{\theta}(a_{i,t}|s_{i,t}))$ to buffer $\mathcal{R}$;\\
					}
					\If{$v_{i}$ collision or pass}
					{
						Exclude $v_{i}$ from $\mathcal{H}$;\\
					}
					Change mode of $v_{i}$;\\
				}
			}
		}
		Calculate $R(s_{i,t},a_{i,t})$ of any $(s_{i,t},a_{i,t},r(s_{i,t},a_{i,t}),s_{i,t}^{'},p_{\theta}(a_{i,t}|s_{i,t}))$ in buffer $\mathcal{R}$;\\
		\For{i from 1 to N}
		{
			Randomly sample a mini-batch from buffer $\mathcal{R}$;\\
			Update $\theta$ according to $\theta \gets \theta + \alpha_1 \nabla_{\theta} J_{PPO}(\theta)$;\\
			Update $\omega$ according to $\omega \gets \omega + \alpha_2 \nabla_{\omega} Loss_{V}(\omega)$;\\
		}
	}	
\end{algorithm}

First, the policy function parameter $\theta$ and value function parameter $\omega$ are randomly initialized, and the buffer $\mathcal{R}$ used for storing the dataset is cleared (line 1-2).
Then, the algorithm iterates $epoch_{math}$ times to optimize $\theta$ and $\omega$. 
In each iteration, $\mathcal{R}$ is cleared first. 
Then, data is collected until $\mathcal{R}$ is full. 
When collecting data, the environment is initialized first. 
On the main road, a random density $\rho$ (vehicles/km) is chosen uniformly in $[\rho_{min},\rho_{max}]$.
Then, each car is assigned a random initial speed $v \sim U(v_{min},v_{max})$. 
Initialization operations similar to those on the main road are performed on the merge road. 
Finally, all vehicles in the scenario are added to the set $\mathcal{G}$, and the global time $t=0$ is initialized (line 6-7).

Next, the vehicles drive in the environment and the algorithm collects vehicle data. 
Since there are multiple vehicles in the environment, depending on the vehicle control mode, the control inputs of all vehicles in $\mathcal{G}$ are collected first (line 10-16), and then executed uniformly (line 18-20). 
During the training period, each vehicle can collect real-time information from all vehicles in the environment. 
For vehicles in CACC control mode, the control method is as described in \SubSecRef{1.CACC control}. 
For vehicles in RL control mode, the state collection method is described in \SubSecRef{2.RL control}. 
Then, the state $s$ is input into $\pi_{\theta}(s)$. 
The output of $\pi_{\theta}(s)$ is two beta distributions corresponding to the acceleration and steering angle. 
The probability density of a beta distribution random variable $X$ is determined by two parameters $\{k,l\}$ and can be expressed as $f(x)=P_{\beta}(X;(k,l))$. 
Therefore, the output of $\pi_{\theta}(s)$ is four-dimensional. Assuming that the output of $\pi_{\theta}(s)$ is $O_{\pi}=\{k_{a},l_{a},k_{\delta},l_{\delta}\}$, the distributions of $a$ and $\delta$ can be expressed as:
\begin{equation}
	\begin{split}
		&f(a)=P_{\beta}(a,(sf(k_{a}+1),sf(l_{a}+1))),\\
		&f(\delta)=P_{\beta}(\delta,(sf(k_{\delta}+1),sf(l_{\delta}+1))),
	\end{split}
\end{equation}
where $sf$ is the \emph{softplus function}. 
Then, the action at this time is sampled based on $f(a)$ and $f(\delta)$. 
During the computation of the action, all vehicles in RL control mode are added to the set $\mathcal{H}$ (line 16). 
Finally, the action is executed for each vehicle in $\mathcal{G}$ in turn.

Next, for each vehicle $v$ in the set, if $v$ is in the set $\mathcal{H}$, the reward $r$ and the new state $s^{'}$ are calculated. In addition, if a vehicle collides or exits the environment, it is removed from the set $\mathcal{G}$. Finally, the vehicles are checked to see if they need to switch control modes and perform the mode switch (line 22-31). During the vehicle's journey, vehicles from the main road always use the CACC control mode. Vehicles from the merging lane use the CACC control mode before point P, and then switch to the RL control mode after passing point P for lateral and longitudinal control. Between point P and point O, the vehicle needs to smoothly transition from the acceleration lane to the main road. After point O, the vehicle returns to the CACC control mode.

After the data collection is completed, the discount return $R(s_t,a_t)=\sum_{i=0} \gamma^{i} r(s_{t+i},a_{t+i})$ is calculated for each data $(s_{i,t},a_{i,t},r(s_{i,t},a_{i,t}),s_{i,t}^{'},p_{\theta}(a_{i,t}|s_{i,t}))$ in $\mathcal{R}$ (line 34).
Finally, the parameters $\theta$ and $\omega$ are updated using the data in buffer $\mathcal{R}$. After $\mathcal{R}$ is full, the parameters are updated $N$ times. In each update, a random mini-batch of data is sampled from the buffer, and $\nabla_{\theta}J(\theta)$ and $\nabla_{\omega}Loss_{V}(\omega)$ are calculated and the gradients are used to update $\theta$ and $\omega$ using gradient descent.

\section{Simulation Results}\label{VII.Simulation Results}\label{simulation results}

\begin{figure}[htbp]
	\centering
	\includegraphics[trim=0.5cm 0.5cm 0.5cm 0.5cm, clip, width=\columnwidth]{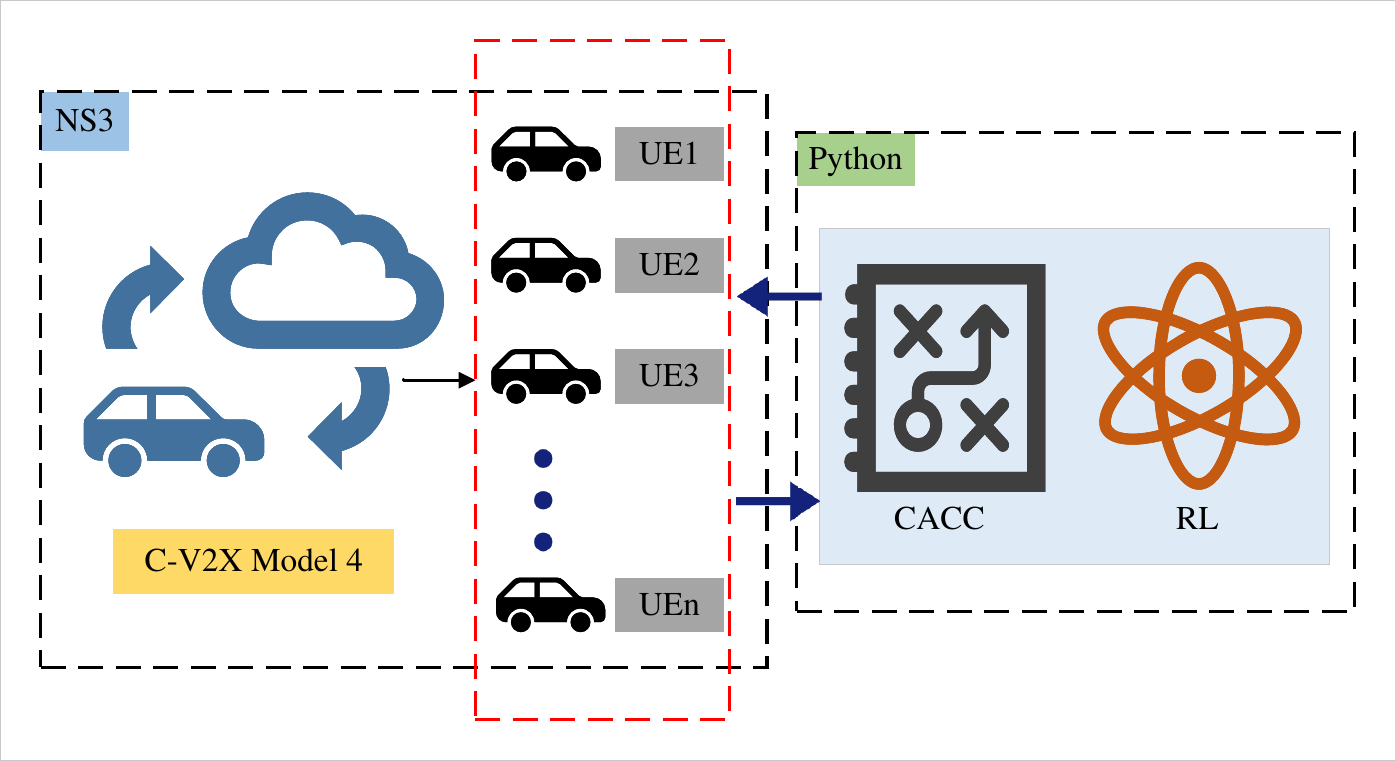}
	\caption{Simulation Platform based on NS3}
	\label{platform}
\end{figure}

This section is divided into two subsections. 
In the first part, we will show the performance improvement of enhanced C-V2X mode 4 to the standard C-V2X mode 4. 
Then, the ramp merging scheme assisted by enhanced C-V2X mode 4 is shown in the second part.
\FigRef{platform} shows the simulation platform we built. 
In this platform, NS3, which is modified from standard version by \Athr{Eckermann} in \cite{Eckermann2019performance}, is responsible for V2X communication simulation and vehicle mobility simulation, while Python is responsible for control. 
Each vehicle adopts the enhanced C-V2X mode 4 to broadcast packets and the communication range of each vehicle is large enough. 
In order to achieve a relatively stable state of communication, the vehicles will communicate for a certain duration before the movement.
\TableRef{v2x-sim-param} shows the V2X protocol parameters used in the simulation.
Moreover, we use the CACC-TP as the comparison scheme, which use CACC and two-point visual control model from \cite{van2021identifiability, doi:10.1068/p5343} as the longitudinal and lateral controller, respectively.

\begin{table}[t]
\centering
\caption{Simulation Parameters of V2X}
\label{v2x-sim-param}
\resizebox{0.45\textwidth}{!}{
\begin{tabular}{|cccc|}
\hline
\multicolumn{4}{|c|}{V2X \& NS3 Parameter}                                                                                              \\ \hline
\multicolumn{1}{|c|}{name}                 & \multicolumn{1}{c|}{value}       & \multicolumn{1}{c|}{name}                  & value      \\ \hline
\multicolumn{1}{|c|}{channel model}        & \multicolumn{1}{c|}{WINNER+B1}   & \multicolumn{1}{c|}{Rsvp}                  & 20         \\ \hline
\multicolumn{1}{|c|}{SC}                   & \multicolumn{1}{c|}{3}           & \multicolumn{1}{c|}{$\beta$}               & 0.0        \\ \hline
\multicolumn{1}{|c|}{subchannel bandwidth} & \multicolumn{1}{c|}{10MHz}       & \multicolumn{1}{c|}{subframe bitmap}       & 0xFFFF     \\ \hline
\multicolumn{1}{|c|}{subchannel scheme}    & \multicolumn{1}{c|}{adjacent}    & \multicolumn{1}{c|}{$T_1$}                 & 4          \\ \hline
\multicolumn{1}{|c|}{$T_2$}                & \multicolumn{1}{c|}{20}          & \multicolumn{1}{c|}{send power}            & 23dBm      \\ \hline
\multicolumn{4}{|c|}{Environment\&Vehicle Parameter}                                                                                    \\ \hline
\multicolumn{1}{|c|}{name}                 & \multicolumn{1}{c|}{value}       & \multicolumn{1}{c|}{name}                  & value      \\ \hline
\multicolumn{1}{|c|}{vehicle length}       & \multicolumn{1}{c|}{4.5m}        & \multicolumn{1}{c|}{vehicle width}         & 2.0m       \\ \hline
\multicolumn{1}{|c|}{$T_h$}                & \multicolumn{1}{c|}{1.0s}        & \multicolumn{1}{c|}{$\Delta T$}            & 100ms      \\ \hline
\multicolumn{1}{|c|}{$v_{min}$}            & \multicolumn{1}{c|}{0m/s}        & \multicolumn{1}{c|}{$v_{max}$}             & 25m/s      \\ \hline
\multicolumn{1}{|c|}{$\rho_{min}$}         & \multicolumn{1}{c|}{$28N_V/km$}  & \multicolumn{1}{c|}{$\rho_{max}$}          & $35N_V/km$ \\ \hline
\multicolumn{1}{|c|}{$a_{min}$}            & \multicolumn{1}{c|}{$-3.0m/s^2$} & \multicolumn{1}{c|}{$a_{max}$}             & $3.0m/s^2$ \\ \hline
\multicolumn{1}{|c|}{$\delta_{min}$}       & \multicolumn{1}{c|}{$-15^\circ$} & \multicolumn{1}{c|}{$\delta_{max}$}        & $15^\circ$ \\ \hline
\multicolumn{1}{|c|}{$v_d$}                & \multicolumn{1}{c|}{20m/s}       & \multicolumn{1}{c|}{$L_a$(adjusting area)} & 200m       \\ \hline
\multicolumn{1}{|c|}{$L_m$(merging area)}  & \multicolumn{1}{c|}{175m}        & \multicolumn{1}{c|}{$R_w$(road width)}     & 3.75m      \\ \hline
\multicolumn{4}{|c|}{CACC param}                                                                                                        \\ \hline
\multicolumn{1}{|c|}{name}                 & \multicolumn{1}{c|}{value}       & \multicolumn{1}{c|}{name}                  & value      \\ \hline
\multicolumn{1}{|c|}{$k_1^1$}              & \multicolumn{1}{c|}{1.0}         & \multicolumn{1}{c|}{$k_2^2$}               & 0.45       \\ \hline
\multicolumn{1}{|c|}{$k_3^2$}              & \multicolumn{1}{c|}{0.125}       & \multicolumn{1}{c|}{$k_2^3$}               & 0.45       \\ \hline
\multicolumn{1}{|c|}{$k_3^3$}              & \multicolumn{1}{c|}{0.05}        & \multicolumn{1}{c|}{$k_2^4$}               & 0.005      \\ \hline
\multicolumn{1}{|c|}{$k_3^4$}              & \multicolumn{1}{c|}{0.05}        & \multicolumn{1}{c|}{}                      &            \\ \hline
\multicolumn{4}{|c|}{PPO Parameters}                                                                                                    \\ \hline
\multicolumn{1}{|c|}{$C_T^1$}              & \multicolumn{1}{c|}{50.0}        & \multicolumn{1}{c|}{$C_T^2$}               & 150.0      \\ \hline
\multicolumn{1}{|c|}{$k_1^r$}              & \multicolumn{1}{c|}{4.3}         & \multicolumn{1}{c|}{$k_2^r$}               & 4.3        \\ \hline
\multicolumn{1}{|c|}{$k_3^r$}              & \multicolumn{1}{c|}{10.0}        & \multicolumn{1}{c|}{$k_4^r$}               & 10.0       \\ \hline
\multicolumn{1}{|c|}{$k_5^r$}              & \multicolumn{1}{c|}{7.5}         & \multicolumn{1}{c|}{$k_6^r$}               & 15.0       \\ \hline
\multicolumn{1}{|c|}{$k_{ego}^x$}          & \multicolumn{1}{c|}{0.05}        & \multicolumn{1}{c|}{$k_{ego}^y$}           & 1.0        \\ \hline
\multicolumn{1}{|c|}{$k_{ego}^\theta$}     & \multicolumn{1}{c|}{1.0}         & \multicolumn{1}{c|}{$k_{ego}^act$}         & 2.0        \\ \hline
\multicolumn{1}{|c|}{$k_\theta^1$}         & \multicolumn{1}{c|}{3.0}         & \multicolumn{1}{c|}{$k_\theta^2$}          & 7.0        \\ \hline
\multicolumn{1}{|c|}{$k_o^{p,1}$}          & \multicolumn{1}{c|}{5.0}         & \multicolumn{1}{c|}{$k_o^{p,2}$}           & 0.7        \\ \hline
\multicolumn{1}{|c|}{$k_d^p$}              & \multicolumn{1}{c|}{5.0}         & \multicolumn{1}{c|}{}                      &            \\ \hline
\end{tabular}
}
\end{table}

\subsection{Simulation Results of Enhanced C-V2X Mode 4}\label{v2x simulation results}

AoI, which is introduced in the first section, is a novel metric to evaluate the performance of network and there have been many work used average AoI in C-V2X mode 4\cite{9380899,parvini2023aoi,9839316,9214855}.
In the scenario of autonomous driving, each vehicle will control itself periodically.
Therefore, for each vehicle, timeliness of information about the vehicles closer to it at the control time is important and it motivates us to propose two novel metrics based on AoI, i.e., \emph{AoI over rate} (AOR) and \emph{position error over rate} (PEOR), which are more suitable than average AoI in the scenario of autonomous driving.

\begin{figure*}[htbp]
    \centering
	\vspace{-1cm}
    \subfigure[0 interference vehicles]{
        \label{AoIOver_0inf}
        \includegraphics[width=0.35\textwidth]{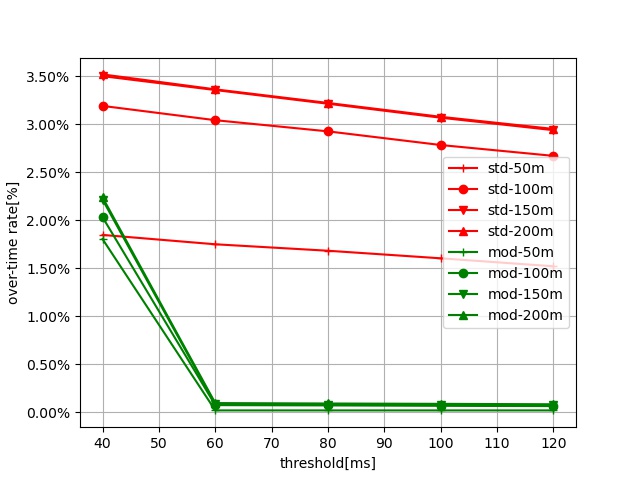}}
	\hspace{-0.9cm} 
    \subfigure[20 interference vehicles]{
        \label{AoIOver_20inf}
        \includegraphics[width=0.35\textwidth]{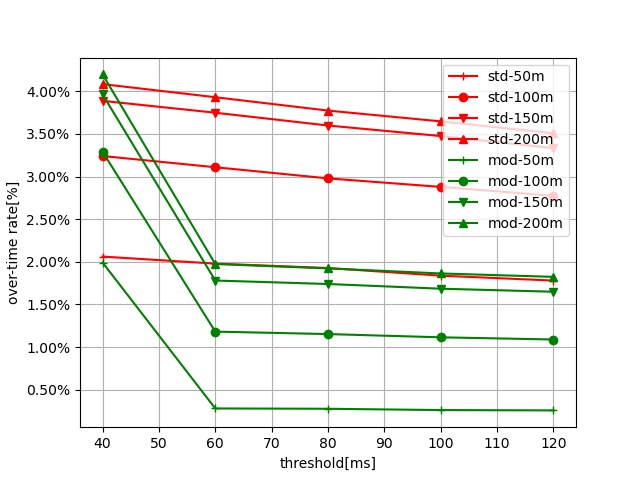}}
	\hspace{-0.9cm} 
    \subfigure[40 interference vehicles]{
        \label{AoIOver_40inf}
        \includegraphics[width=0.35\textwidth]{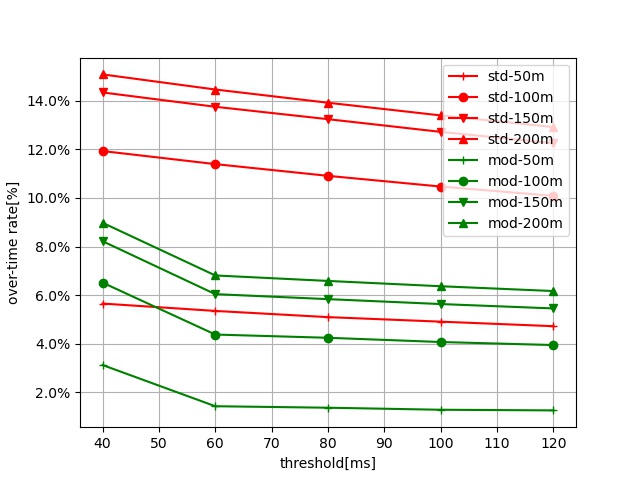}}
    \caption{Comparison of AOR between Enhanced C-V2X Mode 4 and Standard Mode 4 in Different Scenarios}
    \label{AoIOver}
\end{figure*}

\emph{AOR} is define as 
\begin{equation}
\begin{aligned}
	&AoIOver(AoI_{th},d)=\\
	&\frac{\sum_{v \in V} \sum_{t \in C_{v}} \sum_{v' \in V/v} SD(t,v,v',d) SAoI(t,v,v',AoI_{th})}{\sum_{v \in {V}} \sum_{t \in C_{v}} \sum_{v' \in V/v} SD(t,v,v',d)},
\end{aligned}		
\end{equation}
where $V$ is the set of all vehicles in the scenario and $V/v$ is the set of all vehicles except vehicle $v \in G_V$.
Specifically, for each vehicle $v$, we will check AoI of the information received from each vehicle $v'$ in $V/v$ within a distance $d$ at each \emph{control time} and we define all the \emph{control time} of vehicle $v$ as $C_v$. 
At each \emph{control time} $t \in C_v$, the two function, i.e., $SD$ and $SAoI$, will execute.
$SD(t,v,v',d)$ is a binary function and it equals $1$ if the distance of vehicle $v$ and $v'$ is shorter than $d$ at time t, else $0$.
$SAoI(t, v,v',AoI_{th})$ is also a binary function and it equals $1$ if the AoI of $v'$ to $v$ is greater than $AoI_{th}$ and else $0$.
By the definition, we can see that a lower $AOR$ means a higher timeliness of information at \emph{control time} and a better performance in the network.

\emph{PEOR} is similar to \emph{AOR} but it will check the position error rather than AoI.
Specifically, position error is calculated as $\Vert (x_r,y_r) - (x_r,y_r) \Vert_2$, where $(x_r,y_r)$ and $(x_r,y_r)$ represent the coordinate of position obtained by received packet and position at real time.
\begin{equation}
	\begin{aligned}
	&PEOR(e_{th},d)=\\
	&\frac{\sum_{v \in {V}} \sum_{t \in C_{v}} \sum_{v' \in V/v} SD(t,v,v',d) SDist(t,v,v',e_{th})}{\sum_{v \in {V}} \sum_{t \in C_{v}} \sum_{v' \in V/v} SD(t,v,v',d)},
	\end{aligned}
\end{equation}
where $SDist(t,v,v',e_{th})$ is also a binary function and equals $1$ if the position error of vehicle $v'$ to $v$ is greater than the threshold $e_{th}$ and else $0$.
\emph{PEOR} is more directly than \emph{POR} to reflect the quality of service provided by communication at the scenario of autonomous driving when the controller need the assistance of communication.

\begin{figure*}[htbp]
    \centering
    \subfigure[0 interference vehicles]{
        \label{DistOver_0inf}
        \includegraphics[width=0.35\textwidth]{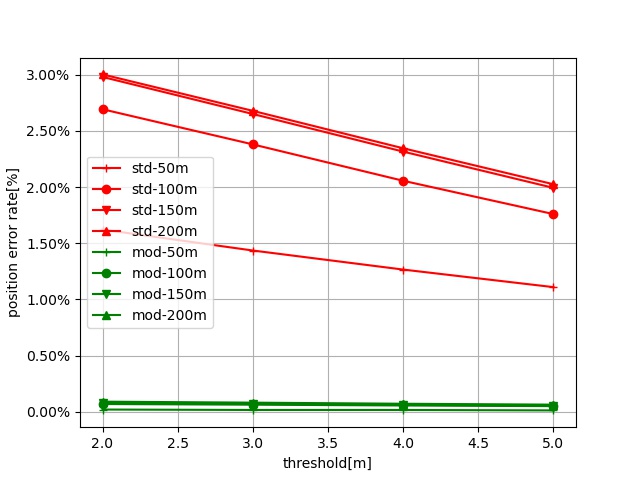}}
	\hspace{-0.9cm} 
    \subfigure[20 interference vehicles]{
        \label{DistOver_20inf}
        \includegraphics[width=0.35\textwidth]{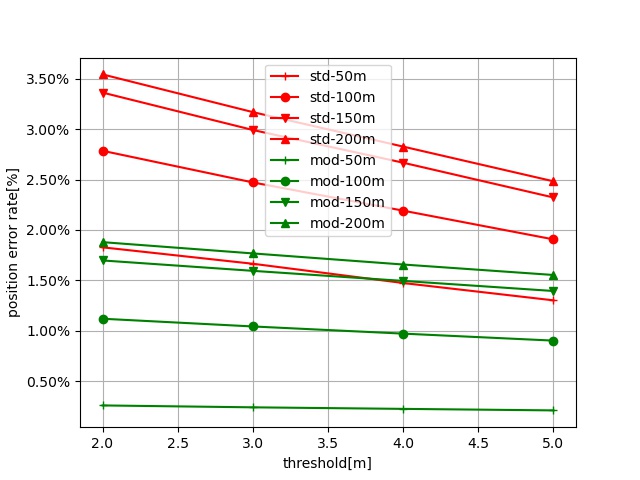}}
	\hspace{-0.9cm} 
    \subfigure[40 interference vehicles]{
        \label{DistOver_40inf}
        \includegraphics[width=0.35\textwidth]{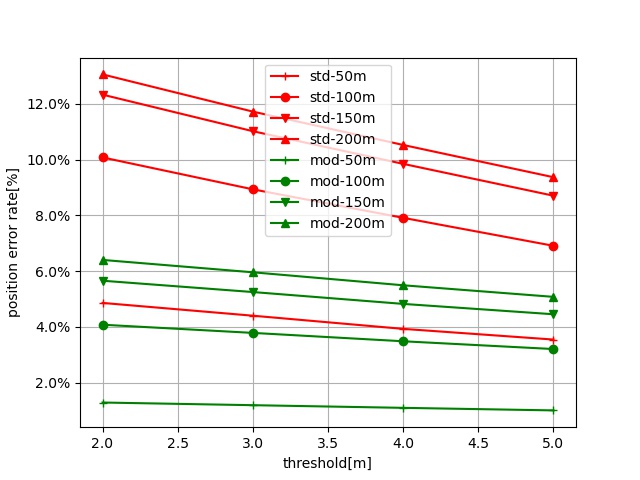}}
    \caption{Comparison of PEOR between Enhanced C-V2X Mode 4 and Standard Mode 4 in Different Scenarios}
    \label{DistOver}
\end{figure*}
\FigRef{AoIOver} shows the simulation results of $AOR(AoI_{th},d)$. 
5 simulations were conducted for each set of parameters, and the average value was calculated. 
In each simulation, the initial position and velocity of the vehicle were initialized by NS3, and the simulation duration was 40 seconds. 
The horizontal axis of the figure is the value of $AoI_{th}$. 
The red and green lines correspond to the standard protocol and the improved protocol proposed in this paper, respectively. 
We also compared the statistical results for different values of $d$. 
From left to right, interference vehicle numbers are 0, 20, and 40. 
We can see that the enhanced C-V2X mode 4 always has lower $AOR$ than standard one.
It is because that the novel resource reservation scheme and SCI format can improve the timeliness of information and the transmission success rate.
In other words, our protocol has an advantage over the original protocol under different communication pressures.
Then, we can see that when there are no interference vehicles, the results of the improved protocol are similar for different values of $d$, while the standard protocol has a large difference between $d=50$ and other values. 
This is because that the enhanced C-V2X mode 4 is not sensitive to changes in $d$ under low communication pressure and can provide better long-distance communication capabilities. 
We can also see that as the number of interference vehicles increases, the overall value of $AOR$ gradually increases. 
This is because that the instantaneous AoI also gradually increases as the communication pressure increases. 

\begin{figure}[htbp]
	\centering
	\includegraphics[scale=0.5]{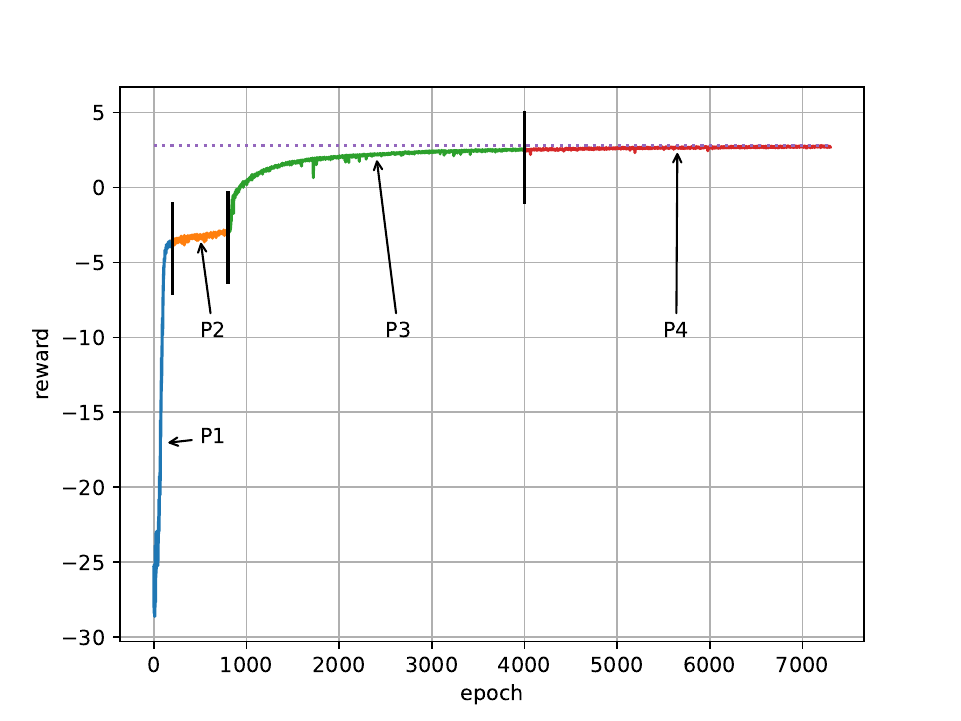}
	\caption{Average Reward during Training Process of PPO}
	\label{average-reward}
\end{figure}

\begin{figure*}[h]
	\centering
	\vspace{-1cm}
	\includegraphics[width=0.8\textwidth]{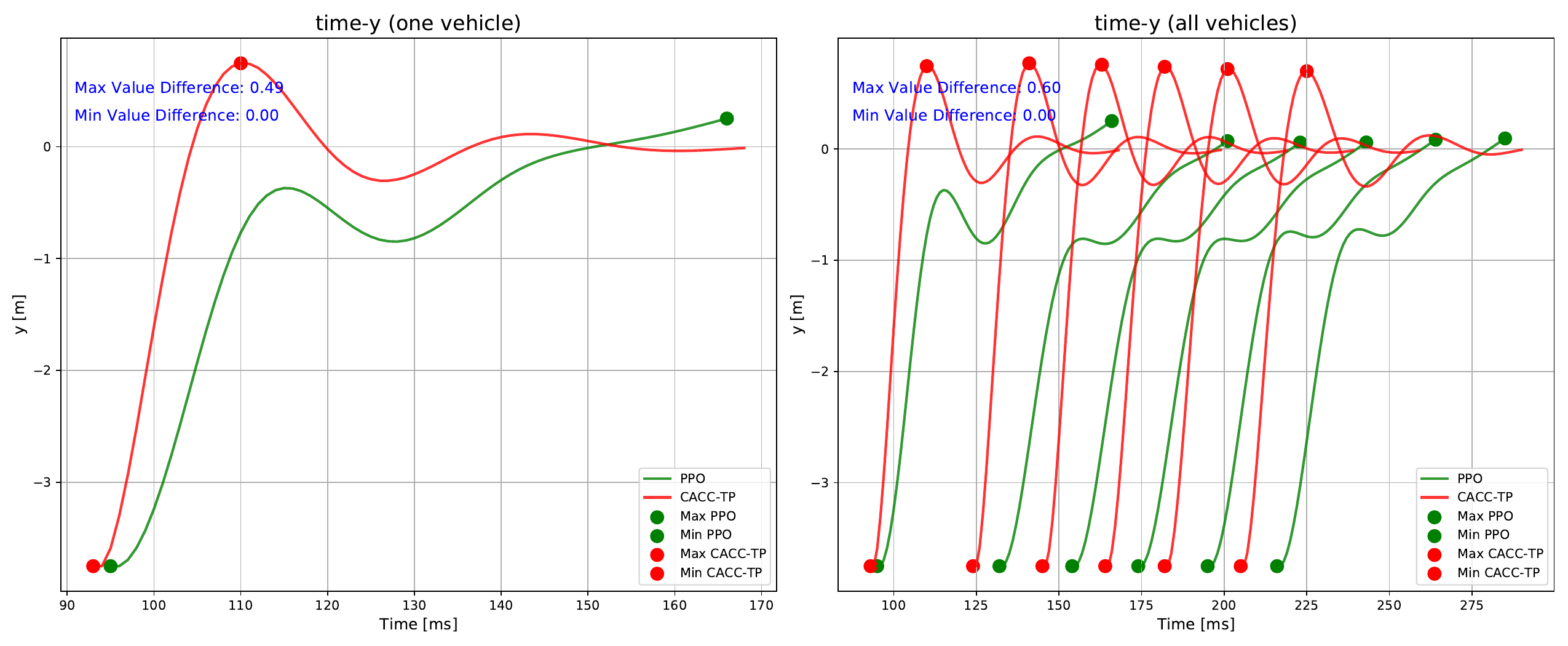}
	\caption{$y$ Coordinates of Vehicles Over Time}
	\label{time-y}
\end{figure*}

\begin{figure*}[h]
	\centering
	\includegraphics[width=0.8\textwidth]{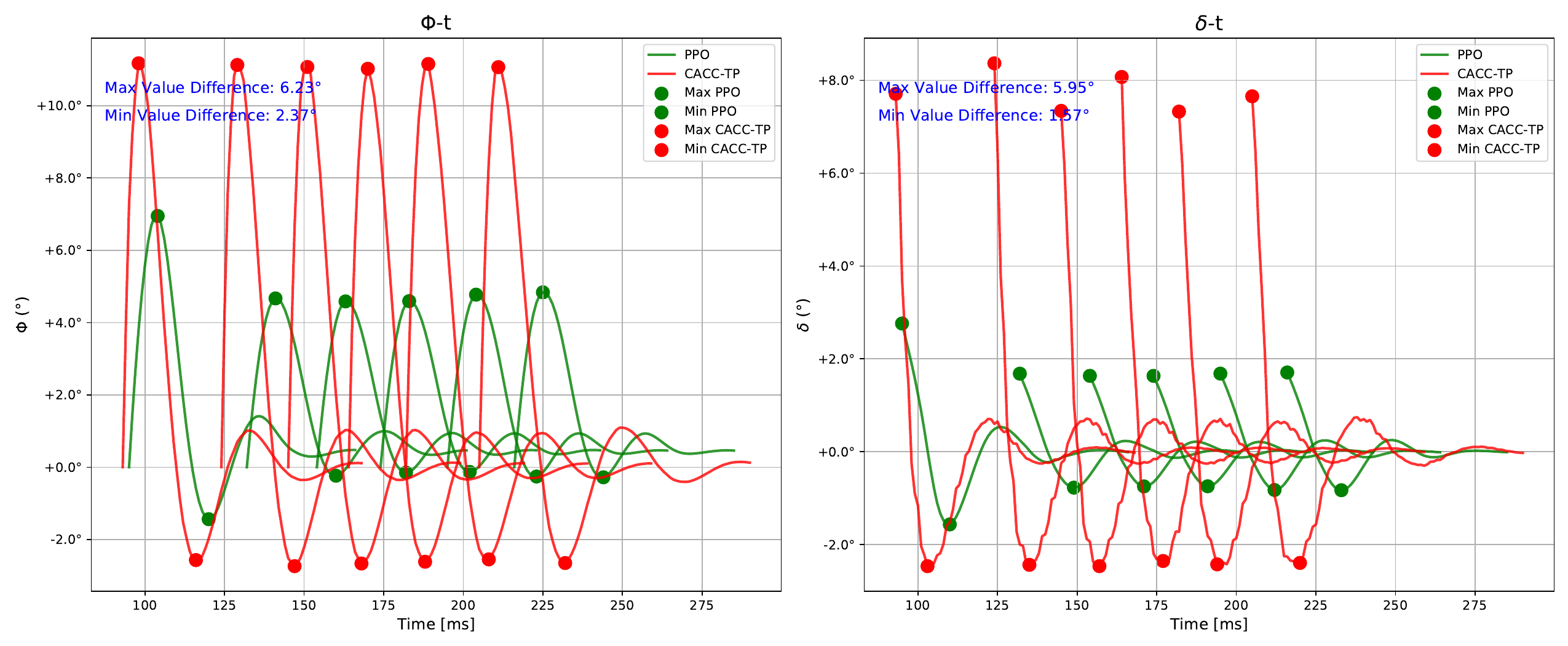}
	\caption{Heading Angle ($\Phi$) and Steering Angle ($\delta$) Over Time}
	\label{time-angle}
\end{figure*}

\subsection{Mobility Simulation Results}\label{vehicle simulation results}

\FigRef{DistOver} shows the comparison between the enhanced C-V2X mode 4 with standard one in terms of $PEOR(d_{th},d)$. 
From left to right are the simulation results with of 0, 20, and 40 interference vehicles. 
The simulation process is the same as that for $PEOR$, with 5 simulations of 40 seconds each and the average result calculated. 
We can see that the enhanced C-V2X mode 4 always has lower PEOR than standard C-V2X mode 4.
It is because the enhanced C-V2X mode 4 has lower AoI at control time, which results in lower position error.
And it indicates that the enhanced C-V2X mode 4 can better performance than standard one.
Moreover, we can see that as the number of interference vehicles increases, the changes of $PEOR$ becomes more pronounced as $d$ changes.
This is because that as the communication pressure increases, the number of resources left in selection window due to $RSRP$ being less than the threshold increases, even though these resources may be occupied by other vehicles. 
In addition, we can see that the difference in $PEOR(d_{th},d)$ is relatively significant at $d_{th}=1.0$ and 2.0. 
It is because that the reason for this phenomenon is similar to that of $PEOR(AoI_{th},d)$. 

The average reward is shown in \FigRef{average-reward}. We can see that the entire training process can be roughly divided into four stages, i.e., P1 to P4.
In P1, average reward value is very low and increases rapidly. In this stage, the vehicles from merging road will not drive along the road, but will collide with other vehicles or the edge of the road, resulting in a lower termination reward, as shown in \EquRef{fail-termination-reward}
As the training progresses, it enters the P2 stage. 
In P2, vehicles will attempt to drive along the acceleration road to avoid collisions with the road edges and vehicles from main road, and terminate at the end of the acceleration lane without merging into the main road.
As shown in \EquRef{fail-termination-reward}, the second item in termination reward, i.e., $-k_1^r \vert x \vert$, is zero, resulting in a higher average reward than P1.
Then, at the end of P2, the vehicles have learn to merge into the main road and enter the P3. 
In P3, the termination reward for vehicles gradually change from \EquRef{success-termination-reward} to \EquRef{fail-termination-reward}.
And the vehicles learn how to control their distance from other vehicles and their posture during the merging process, so the average reward gradually increases.

The figure of the y-axis coordinate of vehicles changing over time is shown in \FigRef{time-y}. 
The $y-t$ of a certain vehicle under different control algorithm is shown in the left subfigure and the right subfigure is the $y-t$ of all vehicles.
Firstly, it can be observed that, compared to the CACC-TP algorithm, our algorithm results in smaller fluctuations in the y-axis of the vehicle, with a tendency toward the negative side of the y-coordinate. This is because our algorithm focuses on the objective described in \EquRef{opt-problem} during the training process, where vehicles merge into the main road from the negative y-coordinate side. The strategy learned by the agent tends to approach the center of the road rather than oscillating around it, whereas the CACC-TP algorithm, influenced by both near and far points during vehicle steering control, causes the vehicle to oscillate around the road center. 
Secondly, we can see that the maximum value of the y-axis in CACC-TP algorithm is about $0.6$m higher than that in our algorithm.
It is because that our algorithm attempts to keep vehicles farther away from the road to keep safety.

The body and steering angle of vehicles from merging road is shown in \FigRef{time-angle}.
In the left subfigure, we can see that the our algorithm's maximum value of a is about 6.23$^\circ$ smaller than algorithm CACC-TP, while the minimum value is about 2.37$^\circ$ smaller. 
That is to say, our algorithm can control $\Phi$ within a small range during the process of ramp merging, thereby providing higher passengers' comfort.
In the right subfigure, we can see that at the beginning of the merging process, CACC-TP algorithm immediately outputs a large steering angle $\delta$, while our algorithm can output a smaller angle.
The difference between the two is approximately 5.95$^\circ$.
It is because that the goal of our algorithm is to maximize value of reinforcement learning, and maximizing value will consider long-term rewards, that is, considering the output of the entire process.
The CACC-TP algorithm outputs a greater value of $\delta$ at beginning, which can immediately bring the vehicle closer to the main road, but it produces a greater value of $\Phi$, requiring a larger negative value of $\delta$, whose absolute value is about 1.57$^\circ$ larger than our algorithm, to be output immediately afterwards.
This will result in higher energy consumption and lower comfort.


\begin{figure}[]
	\centering
	\includegraphics[scale=0.45]{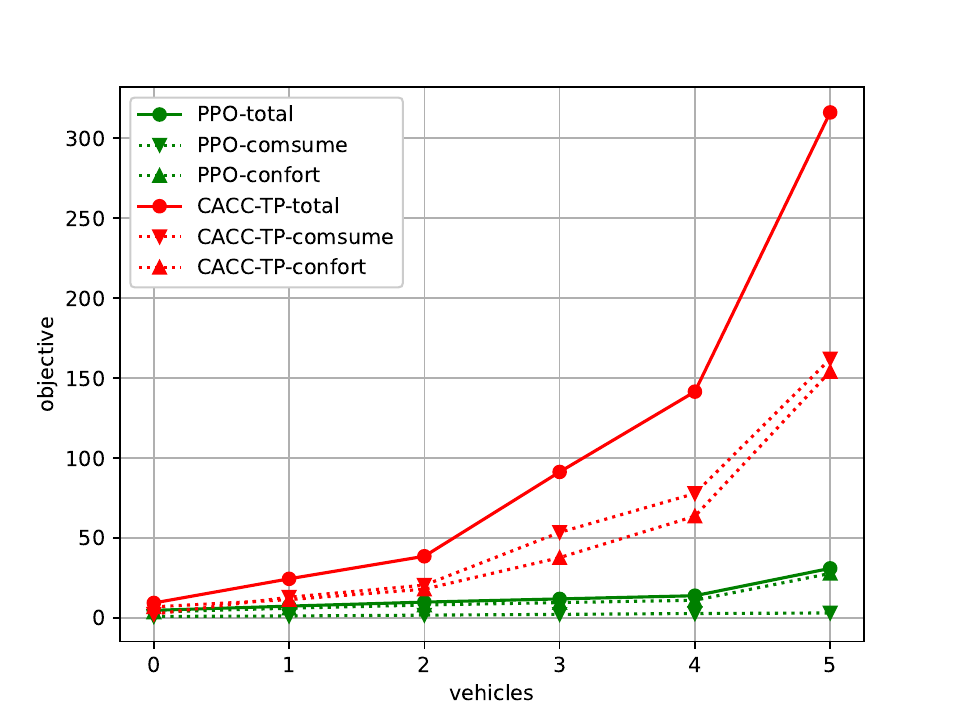}
	\caption{Comparison of Optimization Objectives for Different Algorithm}
	\label{obj}
\end{figure}

\FigRef{obj} shows the objective of in \EquRef{opt-problem} under different vehicle numbers. 
\emph{-total} represents the total objective value, while \emph{consum} and \emph{comfort} represent the proportion of the first item and the proportion of the second item, respectively.
We can see that our algorithm always has lower objective value than CACC-TP algorithm.
It is because in the training process P4, our algorithm learns how to minimize total energy consumption and maximize comfort while maintaining secure inflow. 
In \EquRef{success-termination-reward}, we put these two items into the termination reward to tell the algorithm to learn towards a trajectory which has larger termination reward.
In addition, we can see that with the number of vehicle increasing, the objective gradually increases.
It is because that the number of vehicles increases, the traffic situation becomes more complex. 
In CACC-TP control scheme, once the distance between vehicles approaches a safe distance, vehicles will frequently accelerate and decelerate to maintain the distance between vehicles. 
However, our algorithm considers passengers comfort during the training process, which can achieve smoother acceleration and steering angle changes.

\section{Conclusions}\label{VIII.Concludes}

In this paper, we propose an enhanced C-V2X Mode 4 and a on-ramp merging control scheme which take into account the impact of V2X MAC layer. 
The standard of C-V2X Mode 4 standard has some potential problems which can reduce the timeliness of vehicle control information in autonomous driving scenarios. 
Enhanced C-V2X mode 4 we proposed can improve the timeliness of information and make V2X technology more suitable for autonomous driving scenarios.
With the assistance of V2V technology, we can achieve better traffic performance in the scenario of ramp merging.
In order to achieve better vehicle control, we introduce machine learning into the scenario.
The algorithm based on PPO, a reinforcement learning, can take into account multiple factors, including safety, energy consumption, and comfort. 
The simulation results demonstrated that our control algorithm can optimize the entire integration process, achieving smoother motion trajectories and ultimately enhancing comfort and reducing energy consumption..

\ifCLASSOPTIONcaptionsoff
  \newpage
\fi



\bibliographystyle{template/IEEEtran}
\bibliography{template/refs}

\end{document}